\documentclass[english,global,twocolumn]{svjour}
\usepackage[T1]{fontenc}
\usepackage[latin9]{inputenc}
\usepackage{amsmath}
\usepackage{graphicx}
\usepackage{esint}
\usepackage{babel}

\usepackage[numbers]{natbib}
\begin{document}

\title{Plasmonic nanoparticle monomers and dimers: \\
From nano-antennas to chiral metamaterials}

\author{\author{Dmitry N. Chigrin\inst{1}\and Christian Kremers\inst{1}\and Sergei V. Zhukovsky\inst{1,2}}
\institute{Institute of High-Frequency and Communication Technology, Faculty of Electrical, Information and Media Engineering, University of Wuppertal, Rainer-Gruenter-Str. 21, D-42119 Wuppertal, Germany, \and Department of Physics and Institute for Optical Sciences, University of Toronto, 60 St.George Street, Toronto, Ontario M5S 1A7, Canada}}

\maketitle

\abstract{We review the basic physics behind light interaction with plasmonic
nanoparticles. The theoretical foundations of light scattering on
one metallic particle (a plasmonic monomer) and two interacting particles
(a plasmonic dimer) are systematically investigated. Expressions for
effective particle susceptibility (polarizability) are derived, and
applications of these results to plasmonic nanoantennas are outlined.
In the long-wavelength limit, the effective macroscopic parameters
of an array of plasmonic dimers are calculated. These parameters are
attributable to an effective medium corresponding to a dilute arrangement
of nanoparticles, i.e., a metamaterial where plasmonic monomers or
dimers have the function of {}``meta-atoms''. It is shown that planar
dimers consisting of rod-like particles generally possess elliptical
dichroism and function as atoms for planar chiral metamaterials. The
fabricational simplicity of the proposed rod-dimer geometry can be
used in the design of more cost-effective chiral metamaterials in
the optical domain.}

\section{Introduction}

Light waves cause the charged constituents of matter, electrons and
nuclei, to oscillate. These moving charges in turn emit secondary
light waves, which interfere with the incident wave and with each
other. By an appropriate spatial averaging over the fields at atomic
length scales, the collective response of all constituents can be
derived. This macroscopic response strongly depends on the type of
material. This is how the conventional material equations that relate
the strength and displacement fields in the electromagnetic wave are
introduced, and this is how electrodynamics and optics of homogeneous
media is formulated.

Along similar lines of reasoning, one can engineer the building blocks
({}``meta-atoms'') with feature sizes smaller than the wavelength
of light, and still perform a spatial averaging over the fields at
the {}``meta-atom'' level. This results in {}``macroscopic'' effective
parameters for such artificial composite materials. The ability to
design the meta-atoms in a largely arbitrary fashion adds a new degree
of freedom in material engineering, allowing to create artificial
materials (\emph{metamaterials}) with unusual physical phenomena rare
or absent in nature. Examples include media with negative refractive
index \cite{Mackay2009,MCCALL2008}, hyperbolic or indefinite media
\cite{Hyperbolic2009}, and inhomogeneous media capable of guiding
the radiation around objects ({}``optical cloaking'' \cite{Schurig2006,Cloaking2007,Bilotti2010,Leonhardt2011})
or into objects ({}``optical black holes'' \cite{BlackHole2009})
using the concept of transformation optics \cite{Shalaev2008}.

\begin{figure}
\centering{}\includegraphics[bb=0bp 1bp 170bp 95bp,width=0.95\columnwidth]{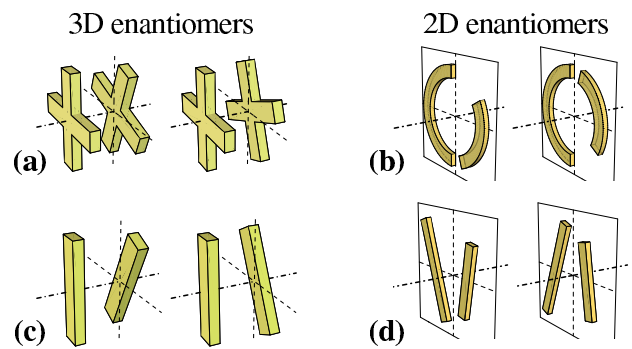}\caption{(Color online) (a--b) Schematic illustration and (c--d) plasmonic
dimer representation of 3D and 2D enantiomeric meta-atoms.\label{fig:enantiomers}}
\end{figure}

Another group of unusual physical phenomena in metamaterials results
from their ability to transform polarization of light. One example
is giant optical activity \cite{1-Giant05} in composite materials
containing spiral-like or otherwise twisted meta-atoms (see, e.g.,
figure~\ref{fig:enantiomers}a). Recently, planar chiral metamaterials
(PCMs) were introduced \cite{POTTS2004,2-Fedotov06,3-Plum09} where
meta-atoms possess two-dimensional (2D) rather than three-dimensional
(3D) enantiomeric asymmetry (figure~\ref{fig:enantiomers}b). While
conventional chiral PCMs simply resemble naturally occurring gyrotopic
media (e.g., optically active liquids), PCMs are distinct from both
3D chiral and Faraday media in that their polarization eigenstates
are co-rotating elliptical rather than counter-rotating elliptical
or circular \cite{2-Fedotov06,3-Plum09,4-Zhukovsky09}. This leads
to exotic polarization properties, e.g., asymmetry in transmission
for a forward- vs.~backward-propagating circularly polarized incident
wave, without nonreciprocity present in Faraday media. Such a rich
variety of polarization properties in compact-sized planar structures
make chiral metamaterials promising for polarization sensitive integrated
optics applications.

Taken generally, the analysis of artificial materials is a complex
theoretical and computational problem. In the theoretical research
of metamaterials, two areas can roughly be identified. On the one
hand, there is the active research concerning various approaches to
metamaterial homogenization, i.e., the derivation of macroscopic material
parameters (in particular, permittivity and permeability tensors)
out of the electromagnetic properties of a single meta-atom \cite{homSimovskyEL,Smith2002,Smith2006,homBaena}.
On the other hand, a lot of effort is invested in the study of light
interaction with single nanoparticles (see \cite{NanoparticleREVIEW}
and references therein). Metallic (plasmonic) nanoparticles attract
special attention \cite{Romero2009} because they can exhibit strong
resonances for the wavelengths much larger than their feature sizes.

Both experimentally and theoretically, one typically takes advantage
of the scalability of Maxwell's equations to consider structures with
millimeter-scale feature sizes in the microwave and radio-frequency
(RF) regime \cite{POTTS2004,2-Fedotov06,3-Plum09}. Not only is the
metamaterial fabrication on these length scales much easier, but also
their theoretical analysis -- meta-atoms have a close resemblance
to antennas in their ability to connect between propagating radiation
and localized near fields. The extensive knowledge in the field of
antenna engineering can then be tapped into. Semi-analytical models
exist \cite{Balanis1989a} that offer insight into field interaction
processes in the antennas, and numerical tools for practical antenna
design and optimization are well established \cite{Volakis2007}. 

The optical wavelength range, which is where functional metamaterials
attract the most interest, remains far more challenging. At optical
frequencies, the metal can no longer be treated as a perfect electric
conductor. Because of this fundmental difference the semi-analytical
models used in the microwave regime can not be used in the visible
and near-infrared spectral range, and the demands in computational
resources increase tremendously. Fabrication of metamaterials in the
optical domain also remains a challenge because of the small feature
sizes (about 10 to 100~nm).

These difficulties notwithstanding, recent progress in nanotechnology
have enabled the fabrication of optical antennas (nanoantennas) \cite{Bharadwaj2009,Muhlschlegel2005}
and opened many exciting possibilities towards nanoantenna applications.
For example, it has been recently demonstrated that nanoantennas can
enhance \cite{Rogobete2007} and direct \cite{Taminiau2008} the emission
of single molecules and that they can play a key role in sensing applications
\cite{Raschke2003}. Great potential in improving the efficiency of
solar cells should also be mentioned \cite{Atwater2010}. 

Such a combination of promises and challenges in optical metamaterials
creates a strong need for a systematic, \emph{ab ibitio}, theoretical
description of metamaterials, which, on the one hand, would be general
enough to span the range of applicable frequencies from RF to optical,
and on the other hand, would be simple enough to facilitate the direct
prediction of metamaterial properties based on their geometry without
having to resort to full-scale numerical simulation in time-consuming
adaptive optimization techniques. Such a theoretical description would
also be beneficial in order to be able to simplify the geometry of
functional (e.g. chiral) meta-atoms so as to facilitate their cost-effective
fabrication in the telecommunication and optical domains. 

In this paper, we formulate such a theoretical description by investigating
the theoretical foundations of light interaction with plasmonic nanoparticles.
We start with one metallic particle (a plasmonic monomer) in an external
electromagnetic field, and consider the general expression for its
susceptibility in terms of the formal Green's function-based solution
of the Maxwell equations. We then set out to simplify this general
solution for a nanoparticle of a simple geometry in the approximation
of its size being much smaller than the wavelength (the Rayleigh approximation),
moving on to the corrections to this approach accounting for the retardation
effects. Further we introduce a new semianalytical method to solve
the problem of electromagnetic wave scattering on metallic cylindrical
wire, which is both accurate and computationally efficient. The results
turn out to be equally applicable in the microwave and optical frequency
range and can be used for improved designs for plasmonic nanoantennas.

We then apply a similar approach to a pair of closely located nanoparticles
(a plasmonic dimer). By accounting for the interaction between the
partices in the dimer, we arrive at analytic expressions of the dimer's
polarizability. It is then possible to determine the effective macroscopic
parameters of a dilute medium that contains the corresponding dimers
as {}``meta-atoms''. It is shown that dimers of planar geometry
without an in-plane mirror symmetry (figure~\ref{fig:enantiomers}d)
exhibit crystallographic and spectral properties of planar chiral
metamaterials. It is concluded that a geometry as simple as an arrangement
of nanorods can support functional metamaterial properties that were
previously reported in more intricately shaped meta-atoms.

The structure of the paper is as follows. In Section \ref{sec:Plasmonic-monomers},
the systematic approach to single plasmonic particles (monomers) is
presented. We start from the general Green's function solution and
review the computational strategies for numerical and semianalytical
approaches, ending up with the expressions of effective particle susceptibility
(polarizability). In Section \ref{sec:Plasmonic-dimers}, a similar
approach is applied to plasmonic dimers. The effective polarizability
of a dimer is determined and the expressions for effective macroscopic
properties such as effective permittivity and permeability is derived.
Application of the developed formalism to 2D planar chiral double-rod
plasmonic dimers is then presented. Finally, Section \ref{sec:Conclusion}
summarizes the paper.

\section{Plasmonic monomers\label{sec:Plasmonic-monomers}}

\subsection{General solution\label{sub:General-solution}}

We describe a nanoparticle by a complex, frequency dependent relative
dielectric permittivity $\varepsilon_{r}$. Electric field $\mathbf{E}\left(\mathbf{r}\right)$
in the presence of such a nanoparticle is given by a solution of the
homogeneous Helmholtz equation \cite{Bladel2007}\begin{equation}
\nabla\times\nabla\times\mathbf{E}\left(\mathbf{r}\right)-k^{2}\varepsilon\left(\mathbf{r}\right)\mathbf{E}\left(\mathbf{r}\right)=0.\label{eq:hom_helmhotz}\end{equation}
Here $k=\omega/c$ is a free space wave number and $\varepsilon\left(\mathbf{r}\right)$
is equal to $\varepsilon_{r}$ inside the volume $V$ occupied by
the particle and to 1 otherwise. Representing the total field $\mathbf{E}\left(\mathbf{r}\right)=\mathbf{E}^{i}\left(\mathbf{r}\right)+\mathbf{E}^{s}\left(\mathbf{r}\right)$
as a sum of incident $\mathbf{E}^{i}\left(\mathbf{r}\right)$ and
scattered $\mathbf{E}^{s}\left(\mathbf{r}\right)$ fields and talking
into account that the incident field has to satisfy a free-space Helmholtz
equation\begin{equation}
\nabla\times\nabla\times\mathbf{E}^{i}\left(\mathbf{r}\right)-k^{2}\mathbf{E}^{i}\left(\mathbf{r}\right)=0\end{equation}
one can reformulate the scattering problem (\ref{eq:hom_helmhotz})
in the form of the inhomogeneous Helmholtz equation \cite{Bladel2007}\begin{equation}
\nabla\times\nabla\times\mathbf{E}^{s}\left(\mathbf{r}\right)-k^{2}\mathbf{E}^{s}\left(\mathbf{r}\right)=\mathrm{i}\omega\mu_{0}\mathbf{j}^{eq}(\mathbf{r})\label{eq:inhom_helmholzu}\end{equation}
with an equivalent current density \begin{equation}
\mathbf{j}^{eq}(\mathbf{r})=-\mathrm{i}\omega\varepsilon_{0}\Delta\varepsilon\left(\mathbf{r}\right)\mathbf{E}(\mathbf{r}),\label{eq:induced_current}\end{equation}
where $\Delta\varepsilon\left(\mathbf{r}\right)=\varepsilon\left(\mathbf{r}\right)-1$.
In this formulation one can clearly see, that the field scattered
from a nanoparticle is generated by a current density induced in the
particle by the total field \textbf{$\mathbf{E}\left(\mathbf{r}\right)$}.

Taking into account the explicit form of the induced current density
(\ref{eq:induced_current}) a general solution of the inhomogeneous
Helmholtz equation (\ref{eq:inhom_helmholzu}) can be written as a
self-consistent integro-differential equation of the Lippman-Schwinger
type \cite{Bladel2007}\begin{multline}
\mathbf{E}(\mathbf{r})=\mathbf{E}^{i}(\mathbf{r})+k^{2}\left(\overleftrightarrow{\mathbf{I}}+\frac{1}{k^{2}}\nabla\otimes\nabla\right)\\
\int_{V}\mathrm{d}^{3}r'g\left(\mathbf{r},\mathbf{r}'\right)\Delta\varepsilon\left(\mathbf{r}'\right)\mathbf{E}\left(\mathbf{r}'\right),\label{eq:lippman-schwinger}\end{multline}
where $\overleftrightarrow{\mathbf{I}}$ denotes the three-dimensional
unit tensor, $\otimes$ is the tensor product defined by $\left(\mathbf{a}\otimes\mathbf{b}\right)_{ij}=a_{i}b_{j}$
and \begin{equation}
g(\mathbf{r},\mathbf{r}')=\frac{e^{\mathrm{i}k\left|\mathbf{r}-\mathbf{r}'\right|}}{4\pi\left|\mathbf{r}-\mathbf{r}'\right|}\label{eq:scalar_greens_function}\end{equation}
is the scalar Green's function. Denoting the integro-differential
operator in (\ref{eq:lippman-schwinger}) by $\overleftrightarrow{\mathbf{t}}$
the Lippman-Schwinger equation can be formally casted in operator
notations as \begin{equation}
\mathbf{E}=\mathbf{E}^{i}+\overleftrightarrow{\mathbf{t}}\mathbf{E}.\end{equation}

To find a general closed-form solution of the Lippman-Schwinger equation
(\ref{eq:lippman-schwinger}) we follow here the approach presented
in \cite{boz2000,boz2001}. It is well known that one can build a
solution of equation (\ref{eq:lippman-schwinger}) in an iterative
manner by replacing a self-consistent field in the integral by the
field obtained on the left-hand side of the equation at the previous
iteration. Starting with the incident field one ends up with an infinite
Born series expansion \cite{Sheng2006}, which in operator notations
reads\begin{multline}
\mathbf{E}=\mathbf{E}^{i}+\left(\overleftrightarrow{\mathbf{t}}+\overleftrightarrow{\mathbf{t}}^{2}+\overleftrightarrow{\mathbf{t}}^{3}\cdots\right)\mathbf{E}^{i}\\
=\mathbf{E}^{i}+\left(\overleftrightarrow{\mathbf{I}}+\overleftrightarrow{\mathbf{t}}+\overleftrightarrow{\mathbf{t}}^{2}+\cdots\right)\overleftrightarrow{\mathbf{t}}\mathbf{E}^{i}\\
=\mathbf{E}^{i}+\left(\overleftrightarrow{\mathbf{I}}-\overleftrightarrow{\mathbf{t}}\right)^{-1}\overleftrightarrow{\mathbf{t}}\mathbf{E}^{i}\label{eq:born-series}\end{multline}
In the last step we have used the formal summation of the operator
power series.

It is important to stress here that the Born expansion of a finite
order can only be used in the case of relatively weak scattering.
In the case of resonant scattering, which is of major importance in
nanoplasmonics, finite-order Born approximation will lead to inaccurate
results no matter how many terms in the expansion are taken into account
\cite{Keller1993}. At the same time, an infinite Born series provides
an exact solution of the Lippman-Schwinger problem (\ref{eq:lippman-schwinger})
both in the resonant and the non-resonant case.

The main idea behind finding an exact solution of (\ref{eq:lippman-schwinger})
is to assume that the scattered field $\mathbf{E}^{s}\left(\mathbf{r}\right)$
can be obtained via some dressed integro-differential operator acting
on the incident field $\mathbf{E}^{i}\left(\mathbf{r}\right)$ directly
\cite{boz2000,boz2001}\begin{multline}
\mathbf{E}^{s}(\mathbf{r})=\mathbf{E}(\mathbf{r})-\mathbf{E}^{i}(\mathbf{r})=k^{2}\left(\overleftrightarrow{\mathbf{I}}+\frac{1}{k^{2}}\nabla\otimes\nabla\right)\\
\int_{V}\mathrm{d}^{3}r'g\left(\mathbf{r},\mathbf{r}'\right)\overleftrightarrow{\Xi}\left(\mathbf{r}'\right)\Delta\varepsilon\left(\mathbf{r}'\right)\mathbf{E}^{i}\left(\mathbf{r}'\right)\label{eq:dressed_lippman}\end{multline}
or in operator notations\begin{equation}
\mathbf{E}=\mathbf{E}^{i}+\overleftrightarrow{\mathbf{T}}\mathbf{E}^{i}.\label{eq:dressed_lippman_operator}\end{equation}
Here the unknown tensor $\overleftrightarrow{\Xi}\left(\mathbf{r}\right)$
has to be determined. Comparing relation (\ref{eq:dressed_lippman_operator})
with the last line of relation (\ref{eq:born-series}) it can be immediately
recognized that the dressed operator $\overleftrightarrow{\mathbf{T}}$
must satisfy the Dyson-like equation

\begin{equation}
\overleftrightarrow{\mathbf{T}}=\overleftrightarrow{\mathbf{t}}+\overleftrightarrow{\mathbf{T}}\overleftrightarrow{\mathbf{t}}.\label{eq:dyson}\end{equation}

To find an unknown tensor $\overleftrightarrow{\Xi}\left(\mathbf{r}\right)$
we act with the operator (\ref{eq:dyson}) on an arbitrary plane harmonic
incident field $\mathbf{E}^{i}\left(\mathbf{r}\right)=\mathbf{E}_{0}\exp\left(\mathrm{i}\mathbf{k}\cdot\mathbf{r}\right)$,
where $\mathbf{k}$ is the wave vector and $\mathbf{E}_{0}$ the amplitude
of the incident field. This leads after a few simple steps to the
following integro-differential equation for the unknown tensor $\overleftrightarrow{\Xi}$\begin{multline}
k^{2}\left(\overleftrightarrow{\mathbf{I}}+\frac{1}{k^{2}}\nabla\otimes\nabla\right)\int_{V}\mathrm{d}^{3}r'\\
\biggl\{\left(\overleftrightarrow{\Xi}\left(\mathbf{r}'\right)g\left(\mathbf{r},\mathbf{r}'\right)\Delta\varepsilon\left(\mathbf{r}'\right)-g\left(\mathbf{r},\mathbf{r}'\right)\Delta\varepsilon\left(\mathbf{r}'\right)\right)\mathrm{e}^{\mathrm{i}\mathbf{k}\cdot\mathbf{r}'}-\\
\overleftrightarrow{\Xi}\left(\mathbf{r}'\right)g\left(\mathbf{r},\mathbf{r}'\right)\Delta\varepsilon\left(\mathbf{r}'\right)k^{2}\left(\overleftrightarrow{\mathbf{I}}+\frac{1}{k^{2}}\nabla'\otimes\nabla'\right)\int_{V}\mathrm{d}^{3}r''\\
g\left(\mathbf{r}',\mathbf{r}''\right)\Delta\varepsilon\left(\mathbf{r}''\right)\mathrm{e}^{\mathrm{i}\mathbf{k}\cdot\mathbf{r}''}\biggr\}\mathbf{E}_{0}=0.\end{multline}

Since the amplitude of the incident field $\mathbf{E}_{0}$ can be
arbitrary, the integral in this equation can be zero only for a vanishing
integrand. That means that the expression in the curly brackets should
be equal to zero leading to\begin{multline}
\overleftrightarrow{\Xi}\left(\mathbf{r}'\right)-\overleftrightarrow{\mathbf{I}}-\overleftrightarrow{\Xi}\left(\mathbf{r}'\right)k^{2}\left(\overleftrightarrow{\mathbf{I}}+\frac{1}{k^{2}}\nabla'\otimes\nabla'\right)\\
\int_{V}\mathrm{d}^{3}r''g\left(\mathbf{r}',\mathbf{r}''\right)\Delta\varepsilon\left(\mathbf{r}''\right)\mathrm{e}^{\mathrm{i}\,\mathbf{k}\cdot\left(\mathbf{r}''-\mathbf{r}'\right)}=0.\end{multline}
This equation can be easily solved to find the unknown tensor\begin{equation}
\overleftrightarrow{\Xi}\left(\mathbf{r}\right)=\left(\overleftrightarrow{\mathbf{I}}-\overleftrightarrow{\Sigma}\left(\mathbf{r}\right)\right)^{-1}\label{eq:tensor-Xi}\end{equation}
 where\begin{multline}
\overleftrightarrow{\Sigma}\left(\mathbf{r}\right)=k^{2}\left(\overleftrightarrow{\mathbf{I}}+\frac{1}{k^{2}}\nabla\otimes\nabla\right)\int_{V}\mathrm{d}^{3}r'\\
g\left(\mathbf{r},\mathbf{r}'\right)\Delta\varepsilon\left(\mathbf{r}'\right)\mathrm{e}^{\mathrm{i}\mathbf{k}\cdot\left(\mathbf{r}'-\mathbf{r}\right)}\label{eq:self-energy}\end{multline}
plays the role of a self-energy operator. With tensor $\overleftrightarrow{\Xi}\left(\mathbf{r}\right)$
found, equation (\ref{eq:dressed_lippman}) gives an exact closed-form
solution of the Lippman-Schwinger equation (\ref{eq:lippman-schwinger}).

It is instructive to reformulate this solution slightly. Recalling
that $\Delta\varepsilon\left(\mathbf{r}\right)=\varepsilon\left(\mathbf{r}\right)-1$
and $\varepsilon\left(\mathbf{r}\right)=\chi\left(\mathbf{r}\right)+1$,
where $\chi\left(\mathbf{r}\right)$ is the electric susceptibility,
one can rewrite the exact solution (\ref{eq:dressed_lippman}) in
the following form\begin{multline}
\mathbf{E}(\mathbf{r})=\mathbf{E}^{i}(\mathbf{r})+k^{2}\left(\overleftrightarrow{\mathbf{I}}+\frac{1}{k^{2}}\nabla\otimes\nabla\right)\int_{V}\mathrm{d}^{3}r'\\
g\left(\mathbf{r},\mathbf{r}'\right)\overleftrightarrow{\mathrm{X}}\left(\mathbf{r}'\right)\mathbf{E}^{i}\left(\mathbf{r}'\right),\label{eq:exact_solution}\end{multline}
where an effective susceptibility tensor of a nanoparticle $\overleftrightarrow{\mathrm{X}}\left(\mathbf{r}\right)=\chi\left(\mathbf{r}\right)\overleftrightarrow{\Xi}\left(\mathbf{r}\right)$
was introduced. Please note that such a general solution is a solution
of the inhomogeneous Helmholtz equation with the equivalent current
density \begin{equation}
\mathbf{j}^{eq}(\mathbf{r})=-\mathrm{i}\omega\varepsilon_{0}\overleftrightarrow{\mathrm{X}}\left(\mathbf{r}\right)\mathbf{E}^{i}(\mathbf{r}),\end{equation}
which justifies the interpretation of $\overleftrightarrow{\mathrm{X}}\left(\mathbf{r}\right)$
as an effective susceptibility tensor connecting an induced linear
polarization $\mathbf{P}\left(\mathbf{r}\right)=\varepsilon_{0}\overleftrightarrow{\mathrm{X}}\left(\mathbf{r}\right)\mathbf{E}^{i}(\mathbf{r})$
of the particle with the incident field $\mathbf{E}^{i}\left(\mathbf{r}\right)$.
In this formulation one can immediately recognize that in the case
of the resonant scattering the integral in (\ref{eq:exact_solution})
will be defined by the poles of the effective susceptibility\begin{equation}
\overleftrightarrow{\mathrm{X}}\left(\mathbf{r}\right)=\chi\left(\mathbf{r}\right)\left(\overleftrightarrow{\mathbf{I}}-\overleftrightarrow{\Sigma}\left(\mathbf{r}\right)\right)^{-1}\label{eq:eff_susceptibility}\end{equation}
as it is required within the linear response theory approach \cite{Abrikosov1965}.

The integrals in (\ref{eq:self-energy}) and (\ref{eq:exact_solution})
are improper at $\mathbf{r}=\mathbf{r}'$ due to a singularity of
the scalar Green's function. Consequently one cannot calculate an
effective susceptibility without an appropriate regularization. To
extract the singularity, one can exclude an arbitrary principal volume
$V^{*}$ around the singular point $\mathbf{r}$, and proceed as in
\cite{Lee1980} to express (\ref{eq:eff_susceptibility}) in the regularized
form

\begin{multline}
\overleftrightarrow{\Sigma}\left(\mathbf{r}\right)=k^{2}\left\{ \int_{V-V^{\star}}\mathrm{d}^{3}r'\overleftrightarrow{\mathbf{G}}(\mathbf{r},\mathbf{r}')\chi\left(\mathbf{r}'\right)\mathrm{e}^{\mathrm{i}\mathbf{k}\cdot\left(\mathbf{r}'-\mathbf{r}\right)}\right.+\\
\int_{V^{\star}}\mathrm{d}^{3}r'\biggl[\overleftrightarrow{\mathbf{G}}(\mathbf{r},\mathbf{r}')\chi\left(\mathbf{r}'\right)\mathrm{e}^{\mathrm{i}\mathbf{k}\cdot\left(\mathbf{r}'-\mathbf{r}\right)}-\\
\frac{1}{k^{2}}\left(\nabla\cdot\left[\nabla g_{0}(\mathbf{r},\mathbf{r}')\right]\right)\chi\left(\mathbf{r}\right)\biggr]\left.-\frac{1}{k^{2}}\overleftrightarrow{\mathbf{L}}_{V^{*}}\chi\left(\mathbf{r}\right)\right\} ,\label{eq:regul_sel-energy}\end{multline}
where $\overleftrightarrow{\mathbf{G}}(\mathbf{r},\mathbf{r}')$ is
the dyadic Green's function\begin{equation}
\overleftrightarrow{\mathbf{G}}(\mathbf{r},\mathbf{r}')=\left(\overleftrightarrow{\mathbf{I}}+\frac{1}{k^{2}}\nabla\otimes\nabla\right)g(\mathbf{r},\mathbf{r}'),\label{eq:dyadic_G}\end{equation}
and $g_{0}$ is the static Green's function\begin{equation}
g_{0}(\mathbf{r},\mathbf{r}')=\lim_{k\rightarrow0}g(\mathbf{r},\mathbf{r}')=\frac{1}{4\pi}\frac{1}{\left|\mathbf{r}-\mathbf{r}'\right|}.\label{eq:static_g0}\end{equation}
$\overleftrightarrow{\mathbf{L}}_{V^{*}}$ denotes the\emph{ }source
dyadic\begin{equation}
\overleftrightarrow{\mathbf{L}}_{V^{*}}=\frac{1}{4\pi}\oint_{\partial V^{\star}}\mathrm{d}^{2}r'\frac{\mathbf{R}\otimes\hat{\mathbf{n}}}{R^{3}}\label{eq:source_dyadic_L}\end{equation}
which accounts for the depolarization of the excluded volume $V^{\star}$
and depends entirely on the geometry of the principal volume, but
does not depend on its size or position \cite{Yaghjian1980}. Here
$\mathbf{R}=\mathbf{r}'-\mathbf{r}$, $R=\left|\mathbf{R}\right|$,
$\partial V^{\star}$ is a surface enclosing principal volume $V^{*}$
and $\hat{\mathbf{n}}$ is its outward unit normal. The dyadic Green's
function defined in (\ref{eq:dyadic_G}) can be calculated explicitly
and is equal to\begin{multline}
\overleftrightarrow{\mathbf{G}}(\mathbf{r},\mathbf{r}')=\frac{e^{\mathrm{i}kR}}{4\pi R}\left\{ \left(1+\frac{\mathrm{i}kR-1}{k^{2}R^{2}}\right)\overleftrightarrow{\mathbf{I}}\right.+\\
\left.\frac{3-3\mathrm{i}kR-k^{2}R^{2}}{k^{2}R^{2}}\frac{\mathbf{R}\otimes\mathbf{R}}{R^{2}}\right\} \label{eq:Green_Dyadic-Free_space}\end{multline}

For points inside the particle , the integral in (\ref{eq:exact_solution})
can be regularized in the similar fashion, resulting in \begin{multline}
\mathbf{E}(\mathbf{r})=\mathbf{E}^{i}(\mathbf{r})+k^{2}\left\{ \int_{V-V^{\star}}\mathrm{d}^{3}r'\overleftrightarrow{\mathbf{G}}(\mathbf{r},\mathbf{r}')\overleftrightarrow{\mathrm{X}}\left(\mathbf{r}'\right)\mathbf{E}^{i}\left(\mathbf{r}'\right)\right.+\\
\int_{V^{\star}}\mathrm{d}^{3}r'\biggl[\overleftrightarrow{\mathbf{G}}(\mathbf{r},\mathbf{r}')\overleftrightarrow{\mathrm{X}}\left(\mathbf{r}'\right)\mathbf{E}^{i}\left(\mathbf{r}'\right)-\\
\frac{1}{k^{2}}\left(\nabla\cdot\left[\nabla g_{0}(\mathbf{r},\mathbf{r}')\right]\right)\overleftrightarrow{\mathrm{X}}\left(\mathbf{r}\right)\mathbf{E}^{i}\left(\mathbf{r}\right)\biggr]\\
\left.-\frac{1}{k^{2}}\overleftrightarrow{\mathbf{L}}_{V^{*}}\overleftrightarrow{\mathrm{X}}\left(\mathbf{r}\right)\mathbf{E}^{i}\left(\mathbf{r}\right)\right\} .\label{eq:regul_solution-2}\end{multline}
For points outside the particle general solution is given by\begin{equation}
\mathbf{E}(\mathbf{r})=\mathbf{E}^{i}(\mathbf{r})+k^{2}\int_{V}\mathrm{d}^{3}r'\overleftrightarrow{\mathbf{G}}(\mathbf{r},\mathbf{r}')\overleftrightarrow{\mathrm{X}}\left(\mathbf{r}'\right)\mathbf{E}^{i}\left(\mathbf{r}'\right).\label{eq:exact_sol_Outside}\end{equation}

\subsection{Rayleigh approximation}

If the nanoparticle dimensions are much smaller than the incident
light wavelength, the scattering problem (\ref{eq:lippman-schwinger})
can be treated within a quasi-static limit $\left|\mathbf{k}\right|=0$.
Letting the principal volume to be infinitesimal, the regularized
integral in (\ref{eq:regul_sel-energy}) can be reduced to \cite{Karam1997}\begin{equation}
\overleftrightarrow{\Sigma}=-\frac{\chi_{r}}{4\pi}\oint_{\partial V}\mathrm{d}^{2}r'\frac{\mathbf{R}\otimes\hat{\mathbf{n}}}{R^{3}}=-\chi_{r}\overleftrightarrow{\mathbf{L}}.\label{eq:self-energy_Rayleigh}\end{equation}
Here the susceptibility of the particle $\chi_{r}$ is taken outside
of the integral, while it is assumed to be constant inside the particle.
The integration is performed over the particle surface $\partial V$.
In (\ref{eq:self-energy_Rayleigh}) we have introduced the depolarization
tensor of the particle $\overleftrightarrow{\mathbf{L}}$. This tensor
depends solely on the particle shape.

Taking into account (\ref{eq:self-energy_Rayleigh}) one obtains a
quasi-static effective susceptibility of the small particle in the
form\begin{equation}
\overleftrightarrow{\mathrm{X}}=\chi_{r}\left(\overleftrightarrow{\mathbf{I}}+\chi_{r}\overleftrightarrow{\mathbf{L}}\right)^{-1}.\label{eq:rayleigh_chi}\end{equation}
The field induced in the particle is given by the quasi-static limit
of (\ref{eq:regul_solution-2})\begin{equation}
\mathbf{E}(\mathbf{r})=\mathbf{E}^{i}-\overleftrightarrow{\mathbf{L}}\overleftrightarrow{\mathrm{X}}\mathbf{E}^{i},\label{eq:rayleigh_field}\end{equation}
which, taking into account that the depolarization tensor is symmetric
\cite{Yaghjian1980}, can be combined with (\ref{eq:rayleigh_chi})
to obtain the familiar relation for the electric field in the Rayleigh
approximation\begin{equation}
\mathbf{E}(\mathbf{r})=\left(\overleftrightarrow{\mathbf{I}}+\chi_{r}\overleftrightarrow{\mathbf{L}}\right)^{-1}\mathbf{E}^{i}.\label{eq:E-static}\end{equation}

Further we calculate the effective quasi-static susceptibility for
a small metallic nanoparticle. Taking into account that the depolarization
tensor is symmetric, the effective susceptibility (\ref{eq:rayleigh_chi})
can be written in component form as\begin{equation}
\overleftrightarrow{\mathrm{X}}=\frac{\chi_{r}}{1+\chi_{r}L_{ii}}\mathbf{\hat{x}}_{i}\otimes\mathbf{\hat{x}}_{i},\label{eq:chi_coord}\end{equation}
where $\mathbf{\hat{x}}_{i}$ are the unit vectors along the main
particle axes. Assuming for simplicity the Drude model for the metal
\begin{equation}
\chi_{r}=-\frac{\omega_{p}^{2}}{\omega^{2}+\mathrm{i}\gamma\omega},\label{eq:drude_chi}\end{equation}
with $\omega_{p}$ being the plasma frequency and $\gamma$ being
the damping rate, the relation (\ref{eq:chi_coord}) results in\begin{equation}
\overleftrightarrow{\mathrm{X}}=\frac{\omega_{p}^{2}}{\omega_{p}^{2}L_{ii}-\omega^{2}-\mathrm{i}\gamma\omega}\mathbf{\hat{x}}_{i}\otimes\mathbf{\hat{x}}_{i}.\label{eq:chi_oscillator}\end{equation}
Then the polarization of the metallic particle $\mathbf{P}=\varepsilon_{0}\overleftrightarrow{\mathrm{X}}\mathbf{E}^{i}$
for an arbitrary plane harmonic incident field $\mathbf{E}^{i}$ should
satisfy the following algebraic equation\begin{equation}
-\omega^{2}\mathbf{P}-\mathrm{i}\gamma\omega\mathbf{P}+\omega_{p}^{2}L_{ii}\mathbf{P}=\varepsilon_{0}\omega_{p}^{2}\mathbf{\hat{x}}_{i}\otimes\mathbf{\hat{x}}_{i}\mathbf{E}^{i},\end{equation}
which after taking an inverse Fourier transform results in the dynamic
equation for a harmonic, damped, driven oscillator\begin{equation}
\frac{\mathrm{d}^{2}\mathbf{P}}{\mathrm{d}t^{2}}+\gamma\frac{\mathrm{d}\mathbf{P}}{\mathrm{d}t}+\omega_{0}\mathbf{P}=\varepsilon_{0}\omega_{p}^{2}\mathbf{\hat{x}}_{i}\otimes\mathbf{\hat{x}}_{i}\mathbf{E}^{i}\label{eq:harmonic_oscil}\end{equation}
with the resonance frequency $\omega_{0}=\omega_{p}^{2}L_{ii}$ defined
by the particle material and shape. Such collective oscillations of
the electron plasma confined to the metallic particle corresponds
to a particle plasmon-polariton excitation.

\subsection{Retardation\label{sub:Retardation}}

With increasing particle size the retardation effects become important
and to obtain an effective susceptibility of the particle as well
as associated scattered field one has to calculate the integrals in
(\ref{eq:self-energy}) and (\ref{eq:exact_solution}) explicitly.
To do that, one should choose the principal volume carefully. Technically
the form and the size of the principal volume can be arbitrary \cite{Lee1980}.
In order to recover the quasi-static limit it is convenient to choose
the principal volume similar in shape to the considered particle.
Furthermore, the principal volume can be chosen infinitesimally small,
resulting in the following exact solution of the scattering problem
for the points inside the particle:\begin{multline}
\mathbf{E}(\mathbf{r})=\mathbf{E}^{i}(\mathbf{r})+k^{2}\int_{V-V^{\star}}\mathrm{d}^{3}r'\\
\overleftrightarrow{\mathbf{G}}(\mathbf{r},\mathbf{r}')\overleftrightarrow{\mathrm{X}}\left(\mathbf{r}'\right)\mathbf{E}^{i}\left(\mathbf{r}'\right)-\overleftrightarrow{\mathbf{L}}\overleftrightarrow{\mathrm{X}}\left(\mathbf{r}\right)\mathbf{E}^{i}\left(\mathbf{r}\right)\label{eq:exact_Final}\end{multline}
with the self-energy tensor given by\begin{multline}
\overleftrightarrow{\Sigma}\left(\mathbf{r}\right)=k^{2}\int_{V-V^{\star}}\mathrm{d}^{3}r'\\
\overleftrightarrow{\mathbf{G}}(\mathbf{r},\mathbf{r}')\chi\left(\mathbf{r}'\right)\mathrm{e}^{\mathrm{i}\mathbf{k}\cdot\left(\mathbf{r}'-\mathbf{r}\right)}-\overleftrightarrow{\mathbf{L}}\chi\left(\mathbf{r}\right).\label{eq:self-energa_Final}\end{multline}
Note that in (\ref{eq:exact_Final}) and (\ref{eq:self-energa_Final})
$\overleftrightarrow{\mathbf{L}}$ is the depolarization tensor of
the particle and the infinitesimal principal volume $V^{*}$ has to
have the same shape as the particle under consideration.

Denoting the result of the integration in (\ref{eq:self-energa_Final})
by $\overleftrightarrow{\mathbf{D}}\left(\mathbf{r}\right)$ one obtains
the effective susceptibility of the particle\begin{equation}
\overleftrightarrow{\mathrm{X}}\left(\mathbf{r}\right)=\chi_{r}\left(\overleftrightarrow{\mathbf{I}}+\chi_{r}\overleftrightarrow{\mathbf{L}}-\overleftrightarrow{\mathbf{D}}\left(\mathbf{r}\right)\right)^{-1}.\label{eq:chi_retarded}\end{equation}
Spatial and frequency dependence of the effective susceptibility originated
from the field retardation is contained in the dynamic depolarization
tensor $\overleftrightarrow{\mathbf{D}}\left(\mathbf{r}\right)$.

To estimate the main effect of the field retardation on the effective
susceptibility tensor we evaluate the dynamic depolarization tensor.
The following approximations are used to simplify the analysis. First,
we assume that the phase difference due to the incident field is negligible
within the particle, i.e., $\exp\left(\mathrm{i}\mathbf{k}\cdot\left(\mathbf{r}'-\mathbf{r}\right)\right)\approx1$
in (\ref{eq:self-energa_Final}). Second, we use only the first few
terms in the long-wavelength expansion of the dyadic Green's function
in powers of $k$ to calculate the integral

\begin{multline}
k^{2}\overleftrightarrow{\mathbf{G}}(\mathbf{r},\mathbf{r}')\approx\frac{1}{4\pi}\left(-\frac{\overleftrightarrow{\mathbf{I}}}{R^{3}}+3\frac{\mathbf{R}\otimes\mathbf{R}}{R^{5}}\right)+\\
\frac{k^{2}}{8\pi}\left(\frac{\overleftrightarrow{\mathbf{I}}}{R}+\frac{\mathbf{R}\otimes\mathbf{R}}{R^{3}}\right)+\frac{\mathrm{i}k^{3}}{6\pi}\overleftrightarrow{\mathbf{I}}.\label{eq:Green_series}\end{multline}
Finally, we use the mean value theorem to approximate the resulting
integral by its value in the geometrical center of the particle.

For a centrosymmetric particle, e.g., a sphere, a regular cylinder,
or a block, the integral of the first term in (\ref{eq:Green_series})
is equal to zero, the second term results in some symmetric shape-dependent
tensor $k^{2}V\overleftrightarrow{\mathbf{D}}'$, while the integral
over the last term results in $\mathrm{i}k^{3}(V/6\pi)\overleftrightarrow{\mathbf{I}}$.
This leads to the following form of the dynamic depolarization tensor\begin{equation}
\overleftrightarrow{\mathbf{D}}\left(\mathbf{r}\right)=k^{2}V\overleftrightarrow{\mathbf{D}}'+\mathrm{i}k^{3}\frac{V}{6\pi}\overleftrightarrow{\mathbf{I}}.\end{equation}
Taking into account that this dynamic depolarization tensor is symmetric,
the effective susceptibility (\ref{eq:rayleigh_chi}) can be written
in component form as\begin{equation}
\overleftrightarrow{\mathrm{X}}=\frac{\chi_{r}}{1+\chi_{r}L_{ii}-k^{2}VD_{ii}'-\mathrm{i}k^{3}\frac{V}{6\pi}}\mathbf{\hat{x}}_{i}\otimes\mathbf{\hat{x}}_{i},\label{eq:chi_retarded_appr}\end{equation}
which for metallic nanoparticle described by the Drude model (\ref{eq:drude_chi})
leads to the following equation for the induced polarization\begin{multline}
-\omega^{2}\mathbf{P}-\mathrm{i}\gamma\omega\mathbf{P}+\omega_{p}^{2}L_{ii}\mathbf{P}+\omega^{4}\frac{V}{c^{2}}\left(D_{ii}'-\frac{\gamma}{6\pi c}\right)\mathbf{P}+\\
\mathrm{i}\omega^{3}\frac{\gamma VD_{ii}'}{c^{2}}\mathbf{P}+\mathrm{i}\omega^{5}\frac{V}{6\pi c^{3}}\mathbf{P}=\varepsilon_{0}\omega_{p}^{2}\mathbf{\hat{x}}_{i}\otimes\mathbf{\hat{x}}_{i}\mathbf{E}^{i}.\label{eq:anharmonic_oscil_omega}\end{multline}
Taking the inverse Fourier transform results in\begin{multline}
\frac{\mathrm{d}^{2}\mathbf{P}}{\mathrm{d}t^{2}}+\gamma\frac{\mathrm{d}\mathbf{P}}{\mathrm{d}t}+\omega_{0}\mathbf{P}+\\
\gamma_{3}\frac{\mathrm{d}^{3}\mathbf{P}}{\mathrm{d}t^{3}}+\omega_{4}\frac{\mathrm{d}^{4}\mathbf{P}}{\mathrm{d}t^{4}}-\gamma_{5}\frac{\mathrm{d}^{5}\mathbf{P}}{\mathrm{d}t^{5}}=\varepsilon_{0}\omega_{p}^{2}\mathbf{\hat{x}}_{i}\otimes\mathbf{\hat{x}}_{i}\mathbf{E}^{i},\label{eq:anharmonic_oscil}\end{multline}
which describes an anharmonic, damped, driven oscillator. Parameters
$\omega_{0}$, $\omega_{4}$, $\gamma_{3}$ and $\gamma_{5}$ can
be easily deduced from the comparison of equations (\ref{eq:anharmonic_oscil_omega})
and (\ref{eq:anharmonic_oscil}). As can be clearly seen from equation
(\ref{eq:anharmonic_oscil}), retardation leads (i) to anharmonicity
and as a consequence to a shift of the resonance frequency (frequency
of the particle plasmon-polariton) and (ii) to additional radiation
damping, which broaden the resonance itself.

\begin{figure}
\begin{centering}
\includegraphics[width=0.8\columnwidth]{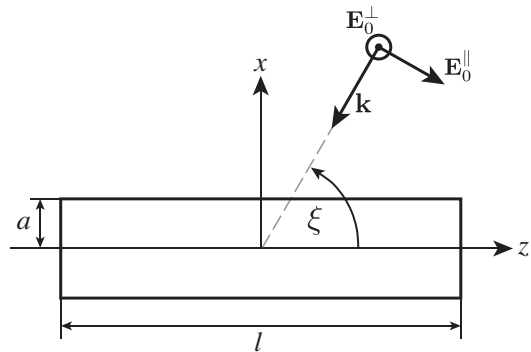} 
\par\end{centering}

\caption{\label{fig:geometry}Definition of the geometrical parameters, radius
$a$ and length $l$, of the scattering cylinder as well as the chosen
body centered coordinate system. Additionally the incident angle $\xi$
and the polarization basis vector $\mathbf{E}_{0}^{\parallel}$ are
depicted in the incident plane.}
\end{figure}

\begin{figure}[t]
\begin{centering}
\includegraphics[width=0.95\columnwidth]{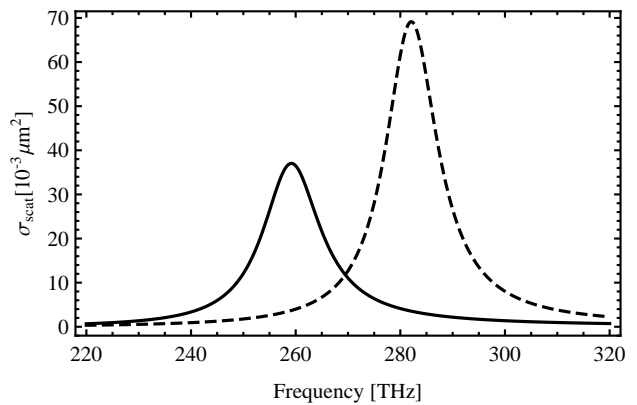}
\par\end{centering}

\caption{Comparison of the scattering cross section for scattering normal incident
light on a gold cylinder (radius 10 nm, length 100 nm) calculated
using the Rayleigh susceptibility (dashed) and the retarded susceptibility
(solid).\label{fig:Comparison-Rayleigh}}
\end{figure}

In figure \ref{fig:Comparison-Rayleigh} the total scattering cross
section for normal incident light scattered on a regular gold nanocylinder
is shown. The geometry of the problem is depicted in figure \ref{fig:geometry}.
The following parameter are used for calculations, particle radius
is $a=10$~nm and its length is $l=100$~nm. The relative permittivity
of gold is modeled by a free electron Drude-Sommerfeld model\begin{equation}
\varepsilon_{r}(\omega)\approx\epsilon_{\infty}-\frac{\omega_{p}^{2}}{\omega^{2}+i\gamma\omega}\label{eq:drude_modell}\end{equation}
with $\varepsilon_{\infty}=9$, $\omega_{p}=1.36674\cdot10^{16}\, s^{-1}$
and $\gamma=7.59297\cdot10^{13}\, s^{-1}$ (see \cite{myro2008}).
Figure \ref{fig:Comparison-Rayleigh} shows both the scattering cross
section obtained from the effective susceptibility (\ref{eq:rayleigh_chi})
within the Rayleigh approximation (dashed line) and from the retarded
effective susceptibility (\ref{eq:chi_retarded_appr}). One can clearly
see that retardation results in a resonance frequency shift towards
lower frequencies and to a broadening of the resonance peak.

\subsection{Numerical solutions}

To obtain a more rigorous solution of the scattering problem, direct
numerical methods have to be used. One class of methods discretize
the volume integral equation, that is equation (\ref{eq:exact_Final})
with the replacement $\overleftrightarrow{\mathrm{X}}\left(\mathbf{r}'\right)\mathbf{E}^{i}\left(\mathbf{r}'\right)\rightarrow\mathbf{E}(\mathbf{r}')$.
A popular representative of this class of solvers is the discrete
dipole approximation (DDA) \cite{Yurkin2007} which solves for induced
dipole moments of small cubes in which the scatterer is decomposed.
A numerical method also based on an integral solution of Maxwell's
equation but expressed in terms of surface instead of volume integrals
is the boundary element method (BEM) \cite{Myroshnychenko2008}. An
advantage of integral equation based methods compared with direct
solutions of Maxwell's equations is that only the scatterer has to
be discretized, in BEM actually just its surface, and one does not
have to deal with unwanted reflections at the borders of the computational
domains. The approach of solving Maxwell's equation in the frequency
domain directly on a volumetric unstructured mesh is called the finite
element method (FEM) \cite{Jin2002}. The idea in FEM is to approximate
the electromagnetic field by polynomials of low order on each small
mesh volume. Plugging this test functions into Maxwell's equation
together with boundary and connection conditions, one yields a system
of linear equations to be solved \cite{Jin2002}. Similar in both
the basic idea and mesh type but formulated in time domain is the
discontinuous Galerkin time domain method (DGTD) \cite{Stannigel2009,Hesthaven2002}.
Another very popular time domain method is the finite-difference time-domain
(FDTD) method \cite{Taflove2000}. Formulated on a structured Cartesian
grid and approximating the spatial and time derivatives by central
differences, this algorithm is explicit, i.e., no set of linear equations
has to be solved. Therefore the implementation of the method is relatively
simple and numerical stability is not a concern as long as space and
time discretization are connected by the Courant relation \cite{Taflove2000}.
On the downside one has to mention that one has to choose a very small
space discretization to accurately describe curved surfaces of scatterers
which leads to high memory consumption and long computation times. 

\begin{figure}
\begin{centering}
\includegraphics[width=0.95\columnwidth]{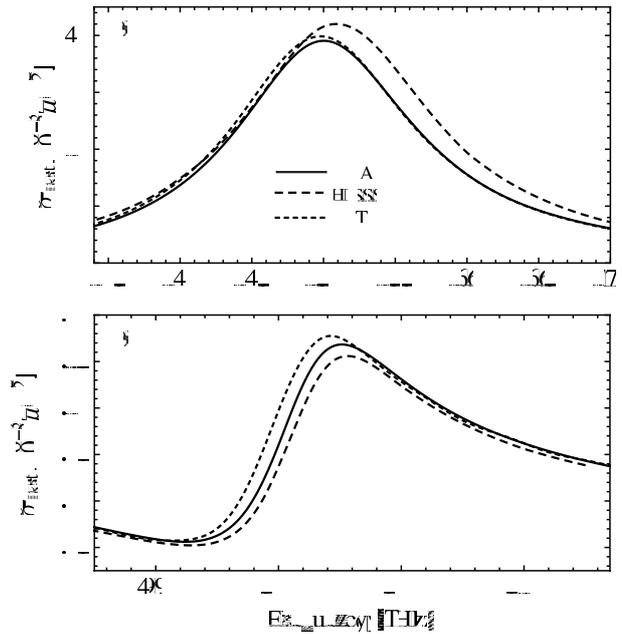} 
\par\end{centering}

\caption{\label{fig:comparison_numeric}Scattering cross-section of a gold
nanowire ($l=200\text{ nm},\, a=10\text{ nm}$) under slanting incidence
($\xi=\frac{\pi}{4}$) calculated with different rigorous numerical
methods. In the top panel (a) the first and in the bottom panel (b)
the third resonance peak are shown.}
\end{figure}

In figure \ref{fig:comparison_numeric} the total scattering cross
section calculated with different direct numerical methods is depicted
for scattering of a plane wave under slanting incidence ($\xi=\frac{\pi}{4}$)
on a gold nanowire with radius $a=10nm$ and length $l=200nm$ (see
figure \ref{fig:geometry}). Gold is described by the equation (\ref{eq:drude_modell})
with the associated parameters. The numerical methods and their implementations
are, in particular: (i) \emph{adda}, an open source DDA implementation
\cite{Yurkin2007a} (solid line); (ii) HFSS, a commercial FEM frequency
domain solver from Ansys Inc.\cite{HFSS} (dashed line); and (iii)
a self-implemented FDTD code (dotted line). The two panels a) and
b) show the first and third resonance peak respectively. The space
discretization in both the DDA and FDTD calculation (both methods
use a cartesian voxel mesh) was set to 1~nm which leads to reasonable
convergence. The mesh in HFSS is created by an adaptive algorithm
so that there is no uniform discretization step. As basis functions
second order polynomials are used. The execution time of DDA and HFSS
per frequency point was about 8 and 6 minutes on one core of a modern
workstation, respectively. The given execution time of HFSS has taken
into account the adaptive recreation of the computational mesh for
each of the depicted panels. FDTD, due to its time domain roots, yields
the whole spectrum in one run which needs about 250 minutes. As it
can be seen in figure \ref{fig:comparison_numeric} the agreement
between the different numerical method are reasonable but not perfect
despite high accuracy in sampling and long execution times.

\subsection{Semi-analytical methods}

\begin{figure}
\begin{centering}
\includegraphics[width=0.95\columnwidth]{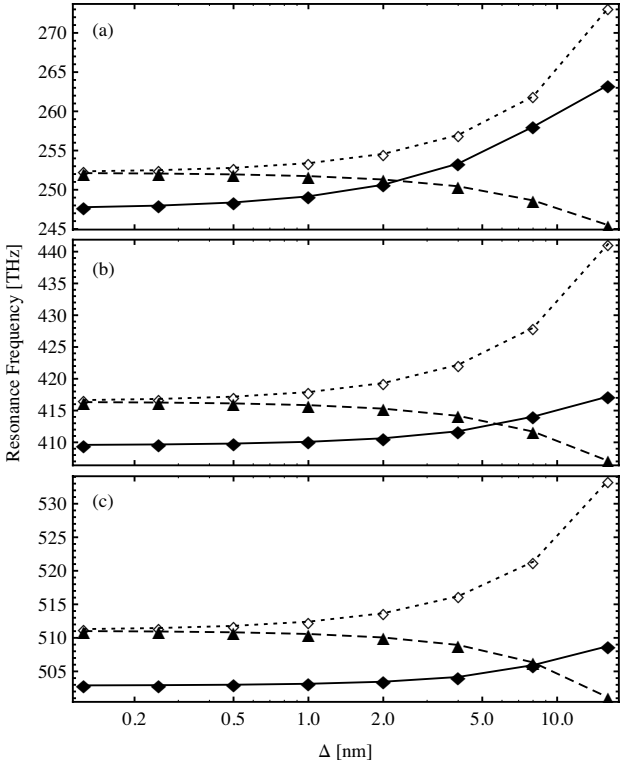} 
\par\end{centering}

\caption{\label{fig:convergence}Convergence of the first three resonances
(from top to bottom panel) of a gold nanowire ($l=200\text{ nm},\, a=10\text{ nm}$)
under slanting incidence ($\xi=\frac{\pi}{4}$) calculated with VC-IE
method (solid line), SI-IE method (dashed) and Hallen's method (dotted).}
\end{figure}
\begin{figure}
\begin{centering}
\includegraphics[width=0.95\columnwidth]{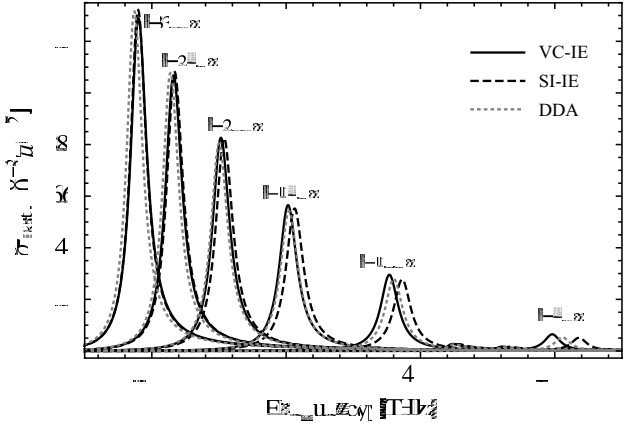} 
\par\end{centering}

\caption{\label{fig:1st_resonances}Scattering cross-sections of gold nanowire
with fixed radius $a=10$ nm and different lengths. The full and dashed
lines show the VC and SI integral equation results, respectively.
The dotted line is the numerically rigorous reference calculated using
a DDA method \cite{Kremers2011}.}
\end{figure}

To overcome the dependence on time consuming numerical simulations
we propose some semi-analytical methods to calculate the scattering
of an incident plane wave on a cylindrical wire as depicted in figure
\ref{fig:geometry}. Further we assume that the radius $a$ of the
wire is small compared with the wavelength of the incident light (electrically
thin wire). In this case the incident light interacting with the wire
can be regarded as a function of $z$ alone\begin{eqnarray}
\mathbf{E}^{i}(\mathbf{r}) & \approx & \mathbf{E}_{0}^{\parallel}e^{-ikz\cos\xi}\label{eq:incident_field-1}\end{eqnarray}
where $\mathbf{E}_{0}^{\parallel}$ denotes the field component along
the wire axis. First we briefly review how the scattering problem
is solved in microwave and RF spectral regime. In this spectral regime
metals can be treated as perfect electric conductor (PEC) with the
consequence that the induced current density has just a longitudinal
$z$-component residing on the wire interface. Expressed in terms
of the total current $I$ the induced current density in this case
can be written as\begin{equation}
\mathbf{j}(\mathbf{r})=\hat{\mathbf{z}}I(z)\frac{\delta(\rho-a)}{2\pi a}\label{eq:surface current density}\end{equation}

By using (\ref{eq:surface current density}) in the Lippman-Schwinger
equation (\ref{eq:lippman-schwinger}) instead of the equivalent current
density (\ref{eq:induced_current}) one derives the integro-differential
equation\begin{multline}
E_{z}(a,z)=E_{z}^{inc}(z)+i\frac{\omega\mu_{0}}{2\pi}\left(1+\frac{1}{k^{2}}\frac{\partial^{2}}{\partial z^{2}}\right)\\
\int_{-\frac{l}{2}}^{\frac{l}{2}}dz'\int_{0}^{2\pi}d\phi'\, g_{a}(\phi',z-z')I(z')\label{eq:pocklington general}\end{multline}
where $g_{a}(\phi',z-z')=g(\mathbf{r},\mathbf{r}')$ with $\mathbf{r}=\left(a,0,z\right)$
and $\mathbf{r}'=\left(a,\phi',z'\right)$ in cylindrical coordinates.
For a perfect electric conductor the tangential electric field at
the wire interface, $E_{z}(a,z)$, vanishes and (\ref{eq:pocklington general})
reduces to the well known Pocklington integro-differential equation
\cite{Orfanidis2010}. In the optical and near-infrared spectral regime
metals exhibit finite conductivity and a skin depth of the order of
typical dimensions of plasmonic particles. However, as long as the
wire is thin, regarding the induced current as pure surface current
should still be a plausible ansatz. In contrast, vanishing tangential
electric field $E_{z}(a,z)$ can not be accepted for wires with finite
conductivity. So one approach to derive a self-consistent equation
to determine $I$ is to express the left hand side of (\ref{eq:pocklington general})
in terms of the current $I$ by utilizing the known exact solution
of the plane wave scattering problem on an infinite cylinder \cite{Bohren1998}.
This should work reasonably well for wires with high aspect ratio,
i.e. $l\gg a$. The analytical solution shows that the electric field
inside an infinitely long, electrically thin cylinder under plane
wave incidence can be approximated with high accuracy by\begin{equation}
\mathbf{E}(\mathbf{r})\approx\hat{\mathbf{z}}f(z)J_{0}\left(k_{\rho}\rho\right)\label{eq:ansatz field}\end{equation}
with $J_{0}$ being the Bessel function of the first kind of 0-th
order and $k_{\rho}=k\sqrt{\varepsilon_{r}-\cos^{2}\xi}$. From this
expression one can derive a relation between the current and the surface
electric field as \begin{equation}
E_{z}(a,z)=Z_{S}I(z)\label{eq:surface impedance relation}\end{equation}
with the surface impedance\begin{equation}
Z_{S}=i\frac{J_{0}(k_{\rho}a)k_{\rho}}{2\pi a\omega\epsilon_{0}\Delta\epsilon_{r}J_{1}(k_{\rho}a)}.\label{eq:surface impedance}\end{equation}
Relation (\ref{eq:surface impedance relation}) can now be used to
express the surface field on the left-hand side of (\ref{eq:pocklington general})
in terms of the current $I$. This yields a self-consistent surface
impedance (SI) integro-differential equation \cite{Hanson2006} \begin{multline}
Z_{S}I(z)=E_{z}^{inc}(z)+i\frac{\omega\mu_{0}}{2\pi}\left(1+\frac{1}{k^{2}}\frac{\partial^{2}}{\partial z^{2}}\right)\\
\int_{-\frac{l}{2}}^{\frac{l}{2}}dz'\int_{0}^{2\pi}d\phi'\, g_{a}(\phi',z-z')I(z')\label{eq:SI-IE}\end{multline}
suitable to calculate the current in thin wires with high aspect ratios
and finite conductivity. However, a more consistent treatment is possible.
One can directly use (\ref{eq:ansatz field}) as an ansatz for the
electric field in (\ref{eq:pocklington general}) on both sides. The
result we obtain is the volume current (VC) integro-differential equation\begin{multline}
f(z)J_{0}\left(k_{\rho}a\right)=E_{z}^{inc}(z)+k^{2}\Delta\epsilon_{r}\left(1+\frac{1}{k^{2}}\frac{\partial^{2}}{\partial z^{2}}\right)\\
\int_{V}d^{3}r'\, g(a,z;\mathbf{r}')f(z')J_{0}\left(k_{\rho}\rho'\right)\label{eq:VC-IE}\end{multline}
which determines the amplitude function $f(z)$. 

Both in equation (\ref{eq:SI-IE}) and in (\ref{eq:VC-IE}) the differential
operator $1+k^{-2}\partial_{z}^{2}$ can be brought inside the integral
by means of Lee's regularization \cite{Lee1980} already used in (\ref{eq:regul_solution-2}).
We thus get rid of the necessity to impose additional boundary conditions.
The regularization step is strictly speaking not needed in case of
the SI integro-differential equation (\ref{eq:SI-IE}) because it
can also be solved in its present form by Hallen's approach \cite{Orfanidis2010}
using the boundary conditions $I(z)=0$ for $z=-l/2$ and $z=l/2$.
However, using pure integral equation is the preferred way because
$I(z)$ is discontinuous at the wire tips so that enforcing the current
to vanish at the wire ends leads to bad convergence when solving (\ref{eq:SI-IE})
numerically. In contrast the VC integro-differential equation can
not be solved in its present form (\ref{eq:VC-IE}) because we do
not know appropriate boundary conditions for the electric field at
the tips. After regularization both the SI-IE and VC-IE can be discretized
within a point matching method of moments (MoM) scheme \cite{Orfanidis2010}.
This leads to an $n$-dimensional matrix equations in the form\begin{equation}
\mathbf{U}=\left(\Gamma\overleftrightarrow{\mathbf{I}}-\overleftrightarrow{\mathbf{M}}\right)^{-1}\mathbf{E}_{z}^{i}\end{equation}
where the dimensionality $n$ is the number of slices in which the
wire is divided. The $i$-th component of $\mathbf{U}$ denotes the
unknown function values $I(z_{i})$ and $f(z_{i})$ at the discrete
point $z_{i}$ for SI-IE and VC-IE respectively. Likewise the $i$-th
component of $\mathbf{E}_{z}^{i}$ denotes the $z$-component of the
incident field at the slice at position $z_{i}$. Obviously, $\Gamma$
and $\overleftrightarrow{\mathbf{M}}$ have to be calculated differently
for SI-IE and VC-IE. In both of them, however, $\Gamma$ contains
the source dyadic and all integration over the slice in which the
singular point resides, whereas the entries of $\overleftrightarrow{\mathbf{M}}$
contain the proper integrals over slices without singularity \cite{Kremers2011}.
The integrations involved in the calculations of $\Gamma$ and the
entries of $\overleftrightarrow{\mathbf{M}}$ as well as the matrix
inversion have to be performed numerically.

\begin{figure*}[t]
\begin{centering}
\includegraphics[width=0.9\textwidth]{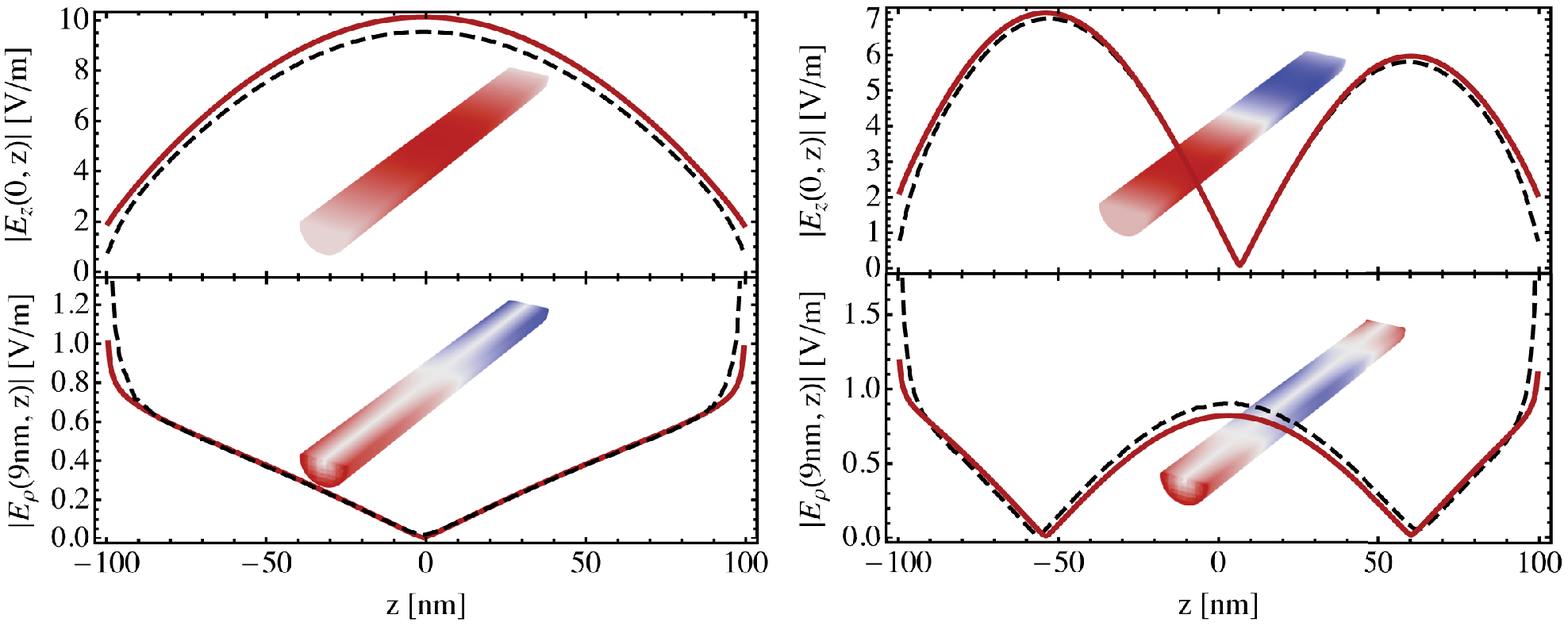} 
\par\end{centering}

\caption{\label{fig:internal field}Induced electric field at the first (left)
and second (right) resonance inside a gold nanowire with radius $a=10$
nm and length $l=200$ nm under slanting incidence ($\xi=\frac{\pi}{4},\, E_{0}^{\parallel}=1$).
Each top panel show the $z$-components on the wire axis and the bottom
panels the radial $\rho$-components 9~nm apart from the axis. The
fields are obtained by HFSS (red solid line) and VC-IE (black dashed
line).}
\end{figure*}

To make sure that an appropriate space discretization $\Delta$ (slice
thickness) is used when applying the VC-IE or SI-IE to realistic scattering
problems, their convergence is studied. Figure \ref{fig:convergence}
shows the dependence of the resonance frequencies on the space discretization
$\Delta$ for a gold nanowire ($l=200\text{ nm},\, a=10\text{ nm}$)
under slanting incidence ($\xi=\frac{\pi}{4}$). From the top to the
bottom panel convergence of resonances of increasing order are shown.
The calculations are performed with (i) the SI-IE solved by Hallen's
approach (dotted line), (ii) the regularized SI-IE (dashed line) and
(iii) the VC-IE (solid line). It can be seen that (i) the converged
result of the SI-IE does not depend on the way of solving it and (ii)
a discretization of $\Delta=1$~nm already shows a deviation from
the converged value of less than 1THz. Thus we choose a discretization
of $\Delta=1$~nm in subsequent studies. In figure \ref{fig:1st_resonances}
the total scattering cross section as function of frequency for gold
nanowires of different lengths but fixed radius, $a=10nm$, under
normal incidence ($\xi=\frac{\pi}{2}$) are shown \cite{Kremers2011}.
The results of VC-IE, SI-IE and numerically rigorous DDA method are
represented by solid, dashed and dotted lines, respectively. One can
trace that (i) the deviation of both semi-analytical methods from
the numerical exact DDA reference solution increase with decreasing
aspect ratios of the wire and (ii) the accuracy of the VC-IE is slightly
better than the one of the SI-IE. This behaviour was expected because
the ansatz (\ref{eq:ansatz field}) for the internal field taken from
the infinitely long wire becomes worse for decreasing aspect ratios.
It is remarkable that for a wire with aspect ratio as low as $\frac{5}{2}$
($l=50nm$) the relative deviation is less than 2\%. But most important
are the differences in execution time. The calculation of one frequency
point in SI and VC integral equation method requires, respectively,
about 1 and 2 seconds on one core of a modern workstation using Mathematica
\cite{mathematica}. An additional advantage of the VC-IE despite
its slightly better accuracy in the calculation of far field properties
like the scattering cross section, is its ability to calculate even
the near field of the plasmonic wire with high accuracy. To demonstrate
this a comparison between the internal field obtained with the VC-IE
and HFSS is shown in figure \ref{fig:internal field} for a gold nanowire
($l=200nm,\, a=10nm$) under slanting incidence ($\xi=\frac{\pi}{4}$).
Here the radial component is calculated by assuming a solution of
the form\begin{equation}
\mathbf{E}(\rho,z)=\hat{\boldsymbol{\rho}}E_{\rho}(\rho,z)+\hat{\mathbf{z}}f(z)J_{0}\left(k_{\rho}\rho\right)\label{eq:internal field form}\end{equation}
and enforcing the field to be divergence free. This results in\begin{equation}
E_{\rho}(\rho,z)=-\frac{J_{1}(k_{\rho}\rho)}{k_{\rho}}\frac{\partial}{\partial z}f(z).\end{equation}
The left and right column of figure \ref{fig:internal field} show
the field components for the first and second resonance of the wire
respectively. The top panel on each column shows the $z$-component
of the electric field at the wire axis and the bottom panel the $\rho$-component
in $9nm$ distance from the axis. The $\rho$-component on the axis
vanishes. The agreement between VC-IE and HFSS is very good. Thus
we can conclude that using the calculated internal field in the form
(\ref{eq:internal field form}), the near field can be calculated
with high accuracy by means of (\ref{eq:exact_sol_Outside}).

\section{Plasmonic dimers\label{sec:Plasmonic-dimers}}

\subsection{Two coupled nanoparticles}

A considerable enhancment of electric field near a resonant plasmonic
nanoparticle opens many application opportunities. Among others, one
should mention strong interaction between the different type of emission
centers and the nanoparticle, which results in strong modification
of emission properties \cite{Kern2011,Giannini2009,Kinkhabwala2009}.
The field enhancement also allows to implement strong coupling among
different nanoparticles themselves. A number of intriguing phenomena
can be observed in coupled plasmonic structures, including Fano-type
resonances \cite{Singh2011,Bachelier2008,Zhang2011}, plasmon-induced
transparency \cite{Zhang2008,Dong2010} (an analog of the electromagnetically
induced transparency \cite{harris1997}) as well as unusual material
properties of metamaterials \cite{Hyperbolic2009,Bilotti2010}.

To rigorously describe even the simplest type of the coupled nanoplasmonic
system, i.e., two particles or a plasmonic dimer, one has to use direct
numerical simulations \cite{Funston2009,Kern2011}. Although a very
intuitive coupled oscillator model \cite{GarridoAlzar2002} can explain
the basic physical mechanisms of the particle plasmon interaction,
it relies on phenomenological fitting parameters and is typically
applicable only to the simplest case of parallel or anti-parallel
particles. In this section we extend the coupled oscillator model
to describe the coupling between two nanoparticles, which can be arbitrarily
oriented with respect to one another. A sketch of the typical structure
under consideration is shown in figure \ref{sec:Plasmonic-dimers}. 

\begin{figure}
\begin{centering}
\includegraphics[width=0.9\columnwidth]{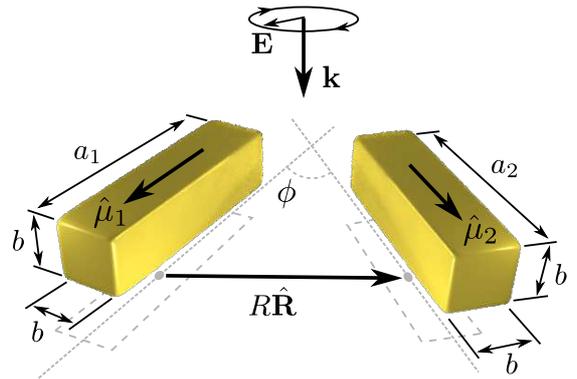}
\par\end{centering}

\caption{Sketch of the dimer structure under consideration. All relevant model
parameters are indicated.\label{fig:dimer}}
\end{figure}

As was demonstrated in Sec.~\ref{sub:General-solution} the effect
of the external field on a nanoparticle can always be described using
an effective susceptibility tensor. Introducing particle polarizability
tensor $\overleftrightarrow{\alpha}_{i}=\varepsilon_{0}\overleftrightarrow{\mathrm{X}}_{i}$
($i=1,\,2$), the polarization (dipole moment) induced in one particle
in presence of the second one can be written in the form\begin{equation}
\tilde{\vec{\mu}}_{i}\left(\mathbf{r}_{i}\right)=\overleftrightarrow{\alpha}_{i}\mathbf{E}\left(\mathbf{r}_{i}\right)\label{eq:dipole_system}\end{equation}
with $\mathbf{E}\left(\mathbf{r}_{i}\right)$ being the total electric
field at the particle $i$ position. Assuming further that electric
dipole response of the particle can be modeled modeled as a response
of the point dipole with the same polarizability the total field $\mathbf{E}\left(\mathbf{r}_{i}\right)$
can be expressed\begin{equation}
\mathbf{E}\left(\mathbf{r}_{i}\right)=\overleftrightarrow{\alpha}_{i}\left(\mathbf{E}_{0}\left(\mathbf{r}_{i}\right)+\frac{k^{2}}{\varepsilon_{0}}\overleftrightarrow{\mathbf{G}}\left(\mathbf{R}\right)\tilde{\vec{\mu}}_{j}\right)\label{eq:dipole_field}\end{equation}
with $\mathbf{R}=\mathbf{r}_{j}-\mathbf{r}_{i}$, where $\mathbf{r}_{i}$
($\mathbf{r}_{j}$) is a position of the $i$th ($j$th) particle.
Here $\overleftrightarrow{\mathbf{G}}$ is the Green's function determined
from equation \eqref{eq:Green_Dyadic-Free_space}, and the second
term is a contribution of the equivalent dipole of the second particle
to the total field at the position of the first particle. To find
an effective induced polarizability of the particles in the dimer,
one has to solve the system (\ref{eq:dipole_system}) with the field
expressed by (\ref{eq:dipole_field}). For example, substituting an
expression of the induce dipole moment of the particle $j$ in the
equation for the dipole moment of the particle $i$ we obtain\begin{multline}
\tilde{\vec{\mu}}_{i}=\overleftrightarrow{\alpha}_{i}\mathbf{E}(\mathbf{r}_{i})=\\
\left(\overleftrightarrow{\alpha}_{i}+\frac{k^{2}}{\varepsilon_{0}}\overleftrightarrow{\alpha}_{i}\overleftrightarrow{\mathbf{G}}(\mathbf{R})\overleftrightarrow{\alpha}_{j}\right)\mathbf{E}_{0}+\\
\frac{k^{4}}{\varepsilon_{0}^{2}}\overleftrightarrow{\alpha}_{i}\overleftrightarrow{\mathbf{G}}(\mathbf{R})\overleftrightarrow{\alpha}_{j}\overleftrightarrow{\mathbf{G}}(-\mathbf{R})\tilde{\vec{\mu}}_{i}\end{multline}
which can be solved to obtain the unknown effective dipole moment
$\tilde{\vec{\mu}}_{i}$\begin{multline}
\tilde{\vec{\mu}}_{i}=\left[\overleftrightarrow{\mathbf{I}}-\frac{k^{4}}{\varepsilon_{0}^{2}}\overleftrightarrow{\alpha}_{i}\overleftrightarrow{\mathbf{G}}(\mathbf{R})\overleftrightarrow{\alpha}_{j}\overleftrightarrow{\mathbf{G}}(-\mathbf{R})\right]^{-1}\\
\left[\overleftrightarrow{\alpha}_{i}+\frac{k^{2}}{\varepsilon_{0}}\overleftrightarrow{\alpha}_{i}\overleftrightarrow{\mathbf{G}}(\mathbf{r})\overleftrightarrow{\alpha}_{j}\right]\mathbf{E}_{0}=\tilde{\overleftrightarrow{\alpha}}_{i}\mathbf{E}_{0}.\label{eq:mu_i}\end{multline}
Here the effective polarizability $\tilde{\overleftrightarrow{\alpha}}_{i}$
of the particle $i$ in the presence of the particle $j$ has been
introduced. Please note that this expression is valid in the point
dipole approximation for an arbitrary mutial orientation of the nanoparticles.

A pair of electrical dipoles is general results in effective electric
dipole, electric quadrupole, and magnetic dipole moments. The effective
dipole moment $\vec{\mu}_{\mathrm{eff}}=\tilde{\vec{\mu}}_{1}+\tilde{\vec{\mu}}_{2}$
is given by a simple sum of the individual dipole moments $\tilde{\vec{\mu}}_{i}$
and results in an effective polarizability of the plasmonic dimer
in the coupled dipole approximation given by\begin{multline}
\overleftrightarrow{\alpha}_{\mathrm{eff}}=\sum_{i=1,2}\left[\overleftrightarrow{\mathbf{I}}-\frac{k^{4}}{\varepsilon_{0}^{2}}\overleftrightarrow{\alpha}_{i}\overleftrightarrow{\mathbf{G}}\overleftrightarrow{\alpha}_{j\neq i}\overleftrightarrow{\mathbf{G}}\right]^{-1}\\
\left[\overleftrightarrow{\alpha}_{i}+\overleftrightarrow{\alpha}_{i}\frac{k^{2}}{\varepsilon_{0}}\overleftrightarrow{\mathbf{G}}\overleftrightarrow{\alpha}_{j\neq i}\right].\label{eq:alpha_eff}\end{multline}
The electric quadrupole and magnetic dipole moments are given respectively
by\begin{equation}
\overleftrightarrow{Q}=\left(\mathbf{r}_{1}\otimes\tilde{\vec{\mu}}_{1}+\tilde{\vec{\mu}}_{1}\otimes\mathbf{r}_{1}+\mathbf{r}_{2}\otimes\tilde{\vec{\mu}}_{2}+\tilde{\vec{\mu}}_{2}\otimes\mathbf{r}_{2}\right)\label{eq:quadrupole}\end{equation}
and\begin{equation}
\mathbf{m}=-\frac{\mathrm{\mathrm{i}\omega}}{2}\left(\mathbf{r}_{1}\times\tilde{\vec{\mu}}_{1}+\mathbf{r}_{2}\times\tilde{\vec{\mu}}_{2}\right).\label{eq:magnetic_dipole}\end{equation}

\begin{figure}
\begin{centering}
\includegraphics[width=0.95\columnwidth]{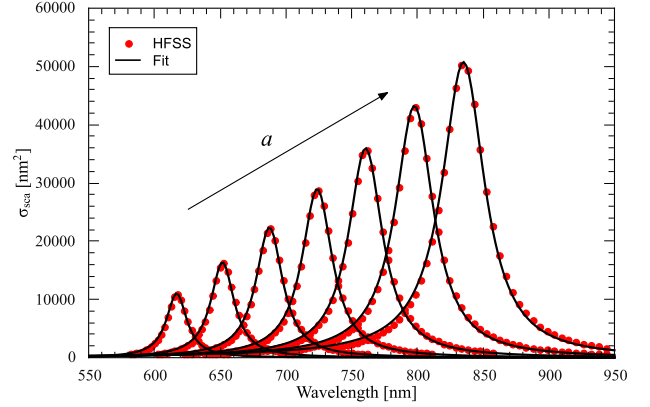}
\par\end{centering}

\caption{Scattering cross sections of single nanoblocks. Lengths are left to
right from 60 nm to 120 nm with 10 nm step. Width and height of all
nanoblocks are 20 nm.\label{fig:single_block}}
\end{figure}

\begin{figure}
\begin{centering}
\includegraphics[width=0.95\columnwidth]{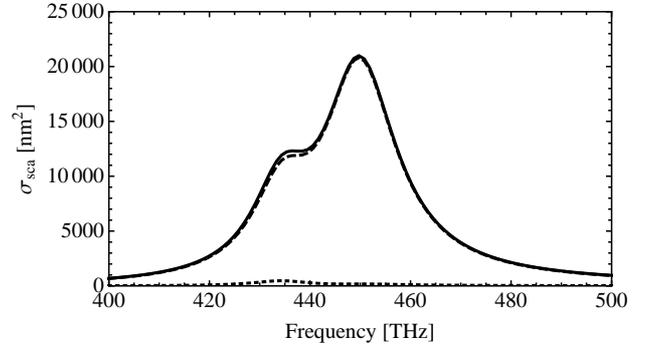}
\par\end{centering}

\caption{Scattering cross section of the gold nanorods dimer for left circular
polarized light calculated within coupled dipole model. Solid line
is a result of contributions from all considered multipole terms.
Dashed line is purely electric dipole contribution. Dotted line shows
combined contribution of magnetic dipole and electric quadrupole.
Gold rods are 100 nm apart and are rotated 45 degree with respect
to each other. Rods are 80 nm and 75 nm long.\label{fig:scs_quadrupole}}
\end{figure}

We further use developed couled dipole model to calculate a scattering
cross section of the gold dimer. For small particles the dominant
contribution to the first resonance is dipole in nature. Moreover
if a particle is elongated, within the spectral range of the longitudinal
plasmon-polariton resonance the contribution of the transverse resonances
can be neglected. With a good approximation corresponding dipole polarizability
can be described by the Lorentz model with all retardation effects
included in the renormalized resonance frequency and damping (Sec.~\ref{sub:Retardation}).
In what follows we model both nanoparticles using dipole polarizability
of the form\begin{equation}
\overleftrightarrow{\alpha}_{j}=\frac{f_{j}\hat{\mu}_{j}\otimes\hat{\mu}_{j}}{\omega_{oj}^{2}-\omega^{2}-i\omega\gamma_{j}}\equiv\alpha_{j}\hat{\mu}_{j}\otimes\hat{\mu}_{j},\label{eq:eq1}\end{equation}
where the resonance frequency $\omega_{oj}$, damping constant $\gamma_{j}$
and amplitude $f_{j}$ are determined by the fit the corresponding
numerical scattering cross section of the individual particles using
the Lorentz model. In figure \ref{fig:single_block} scattering cross
sections of the gold nanoblocks of different length are shown. Both
finite element data (dots) and the results obtained using the fitting
function (\ref{eq:eq1}) are presented, justifying the use of the
proposed approach to modeling of the individual particles.

\begin{figure}
\begin{centering}
\includegraphics[width=1\columnwidth]{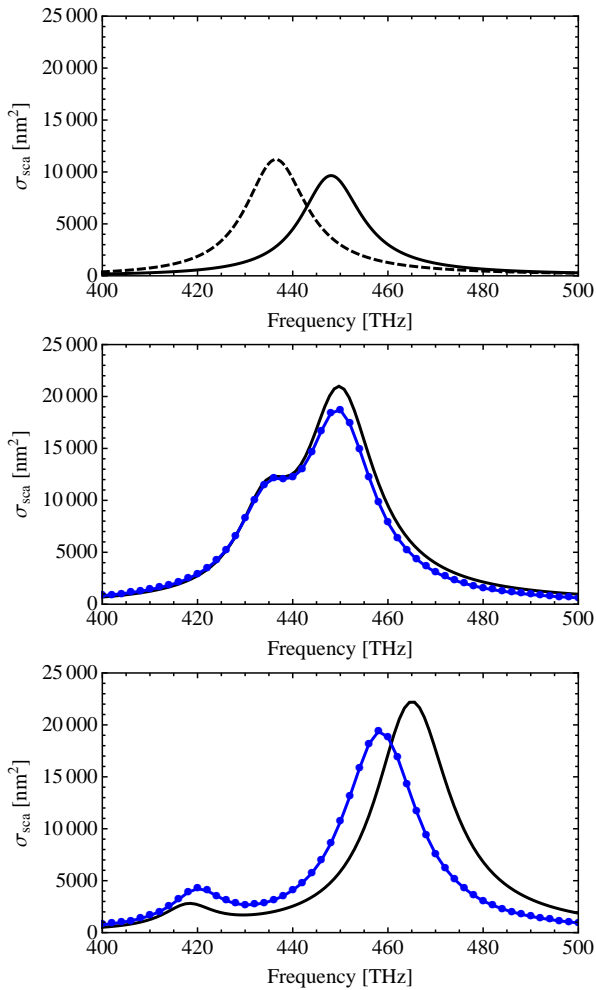}
\par\end{centering}

\caption{Scattering cross section of the gold nanorods dimer for different
rod-to-rod distances. Left circular polarized light is considered.
Distance between rods is 100 nm (middle panel) and 60 nm (bottom panel).
Solid black lines are results of the coupled dipole model. Blue lines
with dots are results of direct numerical simulations performed with
finite element method (HFSS) \cite{HFSS}. In the top panel scattering
cross cross sections are shown for individual rods, 80 nm long (solid
line) and 75 nm long (dashed line). \label{fig:distance}}
\end{figure}

\begin{figure}
\begin{centering}
\includegraphics[width=0.95\columnwidth]{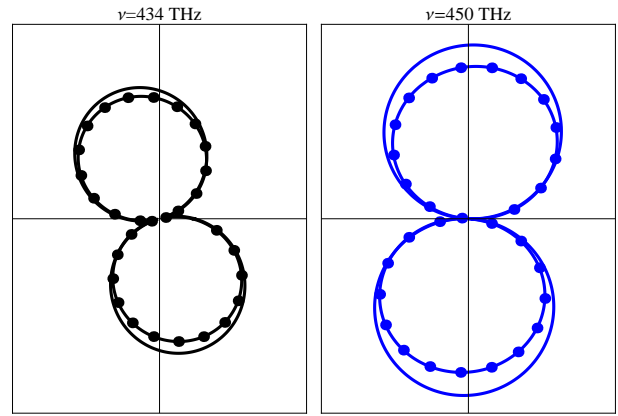}
\par\end{centering}

\caption{Far-field directionality diagrams are shown for two frequencies corresponding
to two maxima in the scattering cross section. Directionality diagrams
are shown in the dimer plane. Solid lines are results of the coupled
dipole model. Lines with dots are results of direct numerical simulations
performed with finite element method (HFSS) \cite{HFSS}. Gold rods
are 100 nm apart and are rotated 45 degrees with respect to each other.
Rods are 80 nm and 75 nm long.\label{fig:directionality-diagrams}}
\end{figure}

In figure \ref{fig:scs_quadrupole} scattering cross section of the
gold dimer build from 80 nm and 75 nm long nanoblocks is shown. Width
and height of both nanoblocks is equal to 20 nm. Particles are situated
100 nm apart and at $45^{\circ}$ with respect to each other. Scattering
cross section is calculated for the left-handed circular polarized
light incident perpendicular to the dimer plane, taking into account
only the electric dipole moment contributions (solid line), the electric
quadrupole and magnetic dipole contributions (dotted line), and all
available multipole contributions (dashed line) . One can clearly
see that for the considered structure the electric dipole contribution
is dominant and contributions from the higher order multipoles can
be neglected.

In figures \ref{fig:distance} and \ref{fig:directionality-diagrams}
comparison of the coupled dipole model (solid lines) with the direct
numerical simulations using finite element method (lines with dots)
is shown. In figure \ref{fig:distance} scattering cross sections
are shown for two distances between particles, namely 100 nm and 60
nm. One can clearly see that with decreasing the distance the quantitative
agreement between analytical and numerical results becomes worse,
but the qualitative agreement remains very reasonable. To appreciate
the amount of the {}``cross-pushing'' (repulsion) of the original
resonances, the scattering cross sections of the individual nanblocks
are presented in the top panel of figure \ref{fig:distance}. Proposed
coupled dipole model demonstrate very good accuracy at moderate particle
separations both for integral and differential scattering properties
as can be seen from figure \ref{fig:directionality-diagrams}, where
a comparison of the analytical and numerical far-field directionality
diagrams is shown.

\subsection{Planar chiral meta-atoms}

Having established that coupled dipole approximation is valis at least
in the case of planar arrangements of nanoblocks in the dimer, we
proceed to derive the effective material parameters. 

We begin by rewriting the equation (\ref{eq:alpha_eff}), using (\ref{eq:eq1}),
in the form\begin{align}
\overleftrightarrow{\alpha}_{\text{eff}}= & \frac{1}{1-\alpha_{1}\alpha_{2}\kappa^{2}}\cdot\biggl[\alpha_{1}\hat{\mu}_{1}\otimes\hat{\mu}_{1}+\alpha_{2}\hat{\mu}_{2}\otimes\hat{\mu}_{2}\nonumber \\
 & +\alpha_{1}\alpha_{2}\kappa\left(\hat{\mu}_{1}\otimes\hat{\mu}_{2}+\hat{\mu}_{2}\otimes\hat{\mu}_{1}\right)\biggr]\nonumber \\
\equiv & \alpha_{1}^{\text{eff}}\hat{\mu}_{1}\otimes\hat{\mu}_{1}+\alpha_{2}^{\text{eff}}\hat{\mu}_{2}\otimes\hat{\mu}_{2}\nonumber \\
 & +\alpha_{3}^{\text{eff}}\left(\hat{\mu}_{1}\otimes\hat{\mu}_{2}+\hat{\mu}_{2}\otimes\hat{\mu}_{1}\right).\label{eq:eq4}\end{align}
 where the unit vectors $\hat{\mu}_{j}$ denote the orientation of
the rods (see figure \ref{fig:dimer}), and$\kappa$ defines the coupling
coefficient between the rods: \begin{equation}
\kappa=\frac{k^{2}}{\varepsilon_{0}}\left[G_{I}\left(\hat{\mu}_{1}\cdot\hat{\mu}_{2}\right)+G_{R}\left(\hat{\mu}_{1}\cdot\hat{\mathbf{R}}\right)\left(\hat{\mu}_{2}\cdot\hat{\mathbf{R}}\right)\right].\label{eq:kappa}\end{equation}

The effective permittivity tensor $\overleftrightarrow{\varepsilon}_{\text{eff}}$
is derived from $\overleftrightarrow{\alpha}_{\text{eff}}$ in equation.~\eqref{eq:eq4}\textbf{
}as \begin{equation}
\overleftrightarrow{\varepsilon}_{\text{eff}}=\overleftrightarrow{\mathrm{I}}-(\varepsilon_{0}V_{\text{cell}})^{-1}\overleftrightarrow{\alpha}_{\text{eff}}.\label{eq:eps_from_alpha}\end{equation}
In axial representation, $\overleftrightarrow{\varepsilon}_{\text{eff}}$
can be expressed as \begin{equation}
\overleftrightarrow{\varepsilon}_{\text{eff}}=\overleftrightarrow{\mathrm{I}}+[\alpha_{1}^{\mathrm{eff}}/(2\epsilon_{0}V_{\mathrm{cell}})]\left(\mathbf{c}_{+}\otimes\mathbf{c}_{-}+\mathbf{c}_{-}\otimes\mathbf{c}_{+}\right)\label{eq:epsilon_eff}\end{equation}
where the complex vectors $\mathbf{c}_{\pm}=\left(\hat{\mu}_{1}+\eta_{\pm}\hat{\mu}_{2}\right)$
determine the directions of the optical axes, and $\eta_{\pm}$ are
given by \begin{equation}
\eta_{\pm}=\frac{\alpha_{3}^{\mathrm{eff}}\pm\sqrt{\left(\alpha_{3}^{\mathrm{eff}}\right)^{2}-\alpha_{1}^{\mathrm{eff}}\alpha_{2}^{\mathrm{eff}}}}{\alpha_{1}^{\mathrm{eff}}}.\label{eq:eta12}\end{equation}

\begin{figure}
\centering{}\includegraphics[width=0.95\columnwidth]{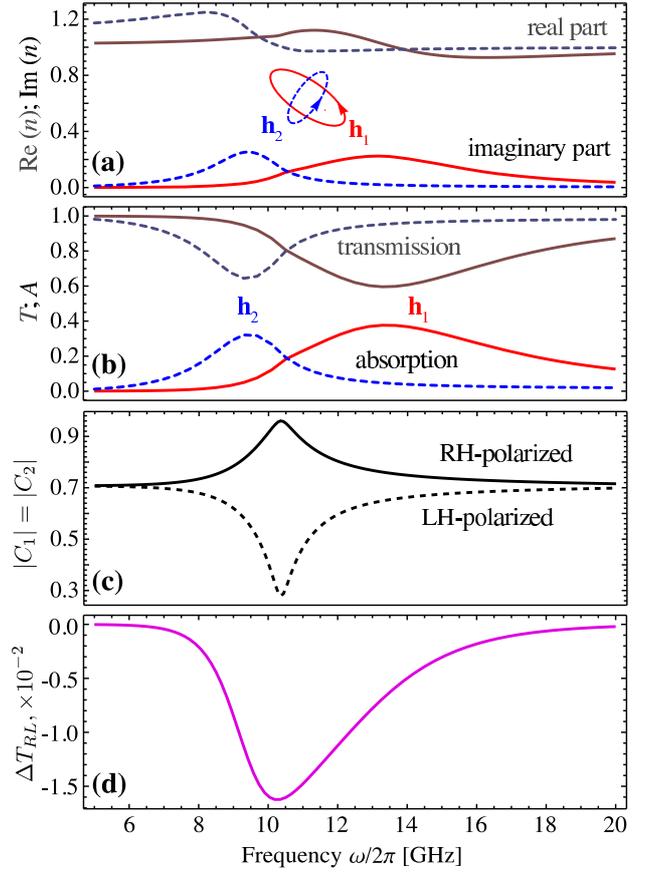}\caption{(Color online) (a) Refractive indices for two co-rotating elliptically
polarized eigenwaves $\mathbf{h}_{1,2}$ of $\protect\overleftrightarrow{N}_{H}$
in equation \eqref{eq:normal_refraction}. (b) Transmission and absorption
for incident wave with polarization given by $\mathbf{h}_{1,2}$.
(c) Absolute value of the complex projection coefficients $C_{1,2}$
{[}see equation \eqref{eq:eigenpolarizations}{]} for RH (solid) and
LH (dashed) circularly polarized incident wave. (d) Analytically calculated
difference in transmission for LH/RH-polarized incident wave. Here
$a_{1}=13$ mm, $a_{2}=10$ mm, $d=10$ mm, $\phi=45^{\circ}$, and
$\psi=0^{\circ}$ \label{fig:dichroism}}
\end{figure}

The dielectric permittivity tensor \eqref{eq:epsilon_eff} corresponds
to an absorbing nonmagnetic crystal, which in general has two distinct
eigenmodes with different polarizations, phase velocities, and absorption
coefficients. These eigenmodes $\mathbf{h}_{1,2}$ and their associated
refractive indices $n_{1,2}$ are given by eigenvectors and eigenvalues
of the refractive index tensor \cite{barkovsky} \begin{equation}
\overleftrightarrow{N}_{H}=\left[\left(-\mathbf{n}^{\times}\overleftrightarrow{\varepsilon}_{\text{eff}}^{-1}\mathbf{n}^{\times}\right)^{-}\right]^{1/2}.\label{eq:normal_refraction}\end{equation}
Here $\mathbf{n}$ is a unit vector of the wave normal. The operator
$\mathbf{n}^{\times}$ acts on a vector $\mathbf{u}$ in such a way
that the result is the vector cross product: $(\mathbf{n}^{\times})\mathbf{u}=\left[\mathbf{n}\times\mathbf{u}\right]$.
The tensor $\overleftrightarrow{A}^{-}$ is defined as a pseudoinverse
to $\overleftrightarrow{A}$ ($\overleftrightarrow{A}^{-}\overleftrightarrow{A}=\overleftrightarrow{\mathrm{I}}-\mathbf{n}\otimes\mathbf{n}$)
and is introduced in place of the true inverse tensor ($\overleftrightarrow{A}^{-1}$),
which does not exist for tensors with zero determinant such as $\mathbf{n}^{\times}$.

In figure \ref{fig:dichroism}a, the real and imaginary parts of the
refractive indices for the two eigenmodes ($n_{1,2}$) are shown for
propagation direction orthogonal to the meta-atoms plane. One can
clearly see that the absorption bands for the two eigenmodes are different.
This is characteristic for dichroic media, and it was shown earlier
that circular or elliptical dichroism is attributable to the optical
manifestations of planar chirality \cite{4-Zhukovsky09}. The eigenvectors
$\mathbf{h}_{1,2}$ turn out to correspond to co-rotating elliptical
polarizations with orthogonal principal axes (see inset in figure~\ref{fig:dichroism}),
which is a signature crystallographic property of planar metamaterials
\cite{2-Fedotov06,3-Plum09}.

Based on the effective permittivity$\overleftrightarrow{\varepsilon}_{\text{eff}}$,
one can determine the optical spectra using the generalized transfer
matrix approach (see, e.g., \cite{6-Borzdov97}). Figure \ref{fig:dichroism}b
shows the transmission and absorption spectra of the medium given
by \eqref{eq:epsilon_eff} for an incident wave with polarization
coincident with the eigenmodes ($\mathbf{h}_{1,2}$). The dips in
the transmission spectra can be seen, in clear correspondence with
the absorption bands. Due to the small variation in the real part
of the refractive indices, reflection from such an effective medium
slab is fairly small and does not contribute to the transmission dips.

For an arbitrarily polarized incident wave, the effective medium acts
as an absorbing polarizing (dichroic) filter splitting the incident
field $\mathbf{H}$ into two waves with polarizations parallel to
the crystal eigenvectors $\mathbf{h}_{1,2}$ as \begin{equation}
\mathbf{H}=C_{1}\mathbf{h}_{1}+C_{2}\mathbf{h}_{2}\label{eq:eigenpolarizations}\end{equation}
where $C_{1,2}$ are complex projection coefficients. The transmission
of the effective medium is then determined, on the one hand, by the
relations between $C_{1,2}$, and on the other hand, on the relations
between the absorption coefficients for the eigenwaves.

Figure \ref{fig:dichroism}c shows the projection coefficients $C_{1,2}$
for circularly polarized incident waves. In this case, it can be seen
that $|C_{1}|=|C_{2}|$. However, in the spectral range of strong
absorption the coupling of RH and LH incident waves to the crystal
eigenmodes $\mathbf{h}_{1,2}$ is very different ($C_{1,2}^{R}\neq C_{1,2}^{L}$).
Hence, circularly polarized waves with different handedness interact
with the metamaterial with a different strength, and consequently,
have different transmittance ($T_{R}\neq T_{L}$). For planar geometry,
reversing the handedness of the incident wave polarization is equivalent
to reversing the direction of incidence or exchange the structure
with its enantiomeric counterpart (see figure \eqref{fig:enantiomers}b,d).
Hence, enantiomeric asymmetry in the transmission $\Delta T_{RL}=T_{R}-T_{L}$
(figure \ref{fig:dichroism}d) can be used to quantify the \textquotedblleft{}strength\textquotedblright{}
of planar chiral properties in a particular structure. It can be seen
that the maximum value of $\Delta T_{RL}$ corresponds to the overlap
between the different absorption bands in the dimer.

\begin{figure*}[t]
\begin{centering}
\includegraphics[width=1\textwidth]{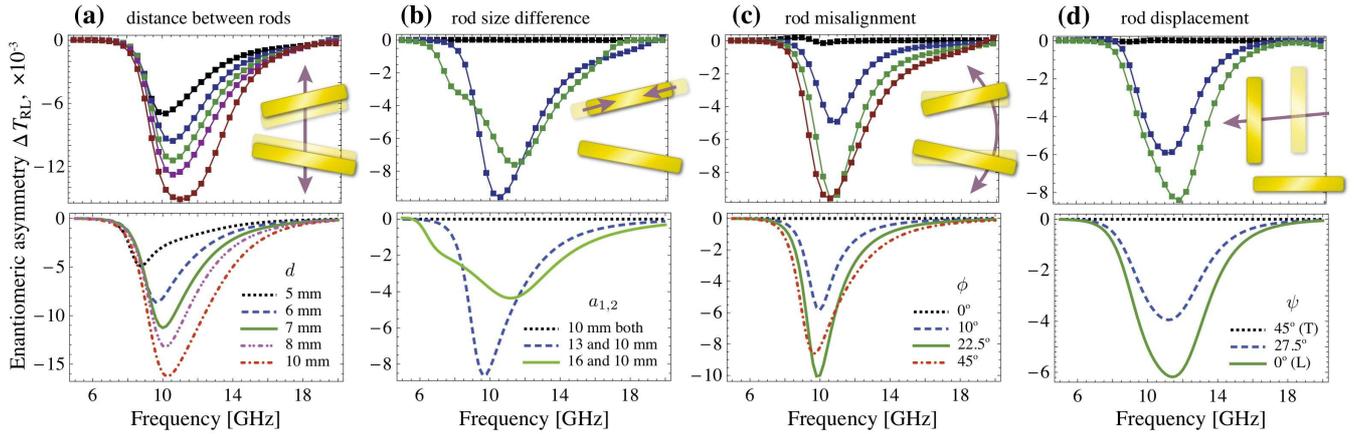}
\par\end{centering}

\caption{(Color online) Numerical (top) and analytical (bottom) dependence
of $\Delta T_{RL}$ on (a) inter-rod distance $d$, (b) difference
in rod length $a_{1}-a_{2}$, (c) rod misalignment angle $\phi$ as
in figure \ref{fig:dimer}, and (c) rod displacement angle $\psi$
for $\phi=90^{\circ}$ (as the dimer changes between T- and L-shaped).\textbf{
}Unless specified otherwise, $a_{1}=13$ mm, $a_{2}=10$ mm, $d=6$
mm, $\phi=45^{\circ}$, and $\psi=0^{\circ}$. \label{fig:results}}
\end{figure*}

Figure \ref{fig:results} shows the spectra $\Delta T_{RL}(\omega)$
for various shapes of the dimer, analytical results from the equations
\eqref{eq:eq1}--\eqref{eq:eq4} compared to the results of direct
3D frequency-domain numerical simulations \cite{HFSS}. It can be
seen that the effective medium model offers a good coincidence with
numerical results.

When the dimer has an in-plane mirror symmetry, the 2D enantiomers
become indistinguishable and such a structure must necessarily be
achiral ($\Delta T_{RL}=0$). For the geometrical transformation considered,
an in-plane mirror symmetry is achieved in the following cases: (i)
for rods of equal length (V-shaped dimer, figure \ref{fig:results}b);
(ii) for parallel rods (II-shaped dimer, figure \ref{fig:results}c);
(iii) for a T-shaped dimer (figure \ref{fig:results}d). In all these
cases, the analytical model correctly predicts the absence of enantiomeric
asymmetry.

Otherwise, the dimers are seen to exhibit chiral properties, which
are stronger when enantiomers are more distinct geometrically. So
there is an optimum rod misalignment angle $\phi\simeq22.5^{\circ}$
in figure \ref{fig:results}c. It is also necessary that the absorption
resonances corresponding to individual eigenpolarizations (figure
\ref{fig:dichroism}c), and hence the resonances of the individual
rods, have some degree of spectral overlap. Hence, $\Delta T_{RL}$
depends on the length mismatch between the rods in a non-monotonic
way: it first increases when the dimer deviates from the achiral V-shape,
reaches a maximum value, and then decreases again with a pronounced
resonance splitting as $\Delta a$ becomes greater (see figure \ref{fig:results}b). 

It can also be noticed that $\Delta T_{RL}$ becomes smaller as the
distance between the rods decreases (figure \ref{fig:results}a).
This can be explained by cross-pushing of the absorption resonances
due to an increase in the inter-rod coupling for smaller inter-rod
distances. The evidence of such cross-pushing can be inferred from
figure \ref{fig:distance}.

If the rods were arranged in a non-planar fashion (e.g., as shown
in figure \ref{fig:enantiomers}c), the magnetic dipole and electric
quadrupole contribution {[}equations (\ref{eq:quadrupole}) and (\ref{eq:magnetic_dipole}){]}
would be expected to contribute to the dimer's electromagnetical response
much more significantly. It can be shown (the detailed calculation
will be available in a forthcoming report) that 3D chiral properties
and giant optical activity effects \cite{1-Giant05} can be expected
from such dimers. This is fully supported by the observation that
some non-planar dimers are geometrically similat to known 3D chiral
meta-atoms (see figures \ref{fig:enantiomers}a,c).

Note, finally, that we have been considering an isolated plasmonic
meta-atom throughout the paper, while real metamaterials contain a
multitude of meta-atoms. Hence, the results obtained here are directly
applicable to metamaterials in the approximation that the meta-atoms
are sufficiently sparse and do not interact with each other. Importantly,
this also means that chiral properties in the metamaterials under
study are intrinsic (attributable to the geometry of the meta-atom
itself) rather than extrinsic (attributable to the meta-atom arrangement,
as in tilted-cross arrays \cite{KseniaPRA09}). This agrees with earlier
time-domain simulation results \cite{7-Kremers09}. With the proposed
model extended to include the inter-atom coupling (see, e.g., \cite{NovitskyArxiv11}),
explicit account of intrinsic vs.~extrinsic effects in 2D and 3D
chiral metamaterials can be given. Combining the intrinsic and extrinsic
contribution in the same metamaterial can be used to maximize its
chiral properties.

\section{Conclusion\label{sec:Conclusion}}

We have systematically described the basic physics of light interaction
with plasmonic nanoparticles. Starting with the general expression
for the polarizability of one metallic particle (a plasmonic monomer)
in an external electromagnetic field, we have explored several ways
to simplify it for the approximation of the nanoparticle size being
much smaller than the wavelength (the Rayleigh approximation). We
have also investigated the corrections to the Rayleigh approach by
accounting for the retardation effects. We have introduced accurate
and numerically efficient 1D semi-analytical method to solve the problem
of light scattering on metallic nanowires for optical frequency range.
We have discussed the applicability of the method and its accuracy
in comparison with the direct numerical simulations. Applications
of these results to plasmonic nanoantennas are outlined.

We have also applied a coupled dipole approach to a pair of closely
located rod-like nanoparticles (a plasmonic dimer). By accounting
for the interaction between the partices in the dimer, we have arrived
at analytic expressions of the dimer's polarizability. The effective
macroscopic material parameters have then been calculated. These parameters
are attributable to an effective medium corresponding to a sparse
arrangement of nanoparticles, i.e., a metamaterial where plasmonic
dimers have the function of {}``meta-atoms''. It is shown that planar
dimers consisting of rod-like particles generally possess elliptical
dichroism \cite{4-Zhukovsky09} and function as atoms for planar chiral
metamaterials\cite{2-Fedotov06,3-Plum09}. It is worth noting that
we have been able to design {}``meta-atom'' structure possessing
the previously reported polarization effects, while being considerably
simpler with respect to analysis and fabrication. Indeed, it is far
easier to fabricate an array of nanorods than it is to fabricate an
array of more complicated objects such as nanogammadions, and the
difference becomes more pronounced as the length scales go down. Hence,
the results of this work can form theoretical grounds to new, cheaper,
easier-to-make chiral metamaterials in the optical domain.

The authors wish to acknowledge financial support from the Deutsche
Forschungsgemeinschaft (DFG) Research Unit FOR 557.

\bibliography{refs}

\end{document}